\documentclass[useAMS,usenatbib]{mn2e}

\usepackage{times}
\usepackage{graphicx,lscape}
\usepackage{fix2col}
\usepackage{graphicx}
\usepackage{ulem}
\usepackage{rotating}
\usepackage{lscape}
\usepackage{longtable}
\usepackage{url}

\def\memsai{Mem.~Soc.~Astron.~Italiana}%

\title[CASE: Deficiency of observed DNe in GCs]
{Cluster AgeS Experiment (CASE): Deficiency of observed \\
dwarf novae in globular clusters}

\author[P. Pietrukowicz et al.]
{P.~Pietrukowicz$^{1,2}$\thanks{E-mail: pietruk@camk.edu.pl},
J.~Kaluzny$^{1}$, A.~Schwarzenberg-Czerny$^{1,3}$,
I.~B.~Thompson$^{4}$, \newauthor W.~Pych$^{1}$, W.~Krzeminski$^{5}$
and B.~Mazur$^{1}$ \\
$^{1}$Nicolaus Copernicus Astronomical Center,
      ul. Bartycka 18, 00-716 Warsaw, Poland \\
$^{2}$Departamento de Astronom\'ia y Astrof\'isica, Pontificia
      Universidad Cat\'olica de Chile, Casilla 306, Santiago 22, Chile \\
$^{3}$Astronomical Observatory, Adam Mickiewicz University,
      ul. S{\l}oneczna 36, 61-286 Pozna\'n, Poland \\
$^{4}$Carnegie Institution of Washington, 813 Santa Barbara Street,
      Pasadena, CA 91101, USA \\
$^{5}$Las Campanas Observatory, Casilla 601, La Serena, Chile \\
}

\begin{document}

\date{Accepted ..... Received .....; in original form .....}

\pagerange{\pageref{firstpage}--\pageref{lastpage}} \pubyear{2008}

\maketitle

\label{firstpage}

\begin{abstract}
We present the results of a search for dwarf novae (DNe) in globular
clusters (GCs). It is based on the largest available homogeneous sample of
observations, in terms of the time span, number of observations
and number of clusters. It includes 16 Galactic GCs and yielded
two new certain DNe: M55-CV1 and M22-CV2. All previously known
systems located in our fields were recovered, too. We surveyed M4,
M5, M10, M12, M22, M30, M55, NGC 288, NGC 362, NGC 2808, NGC 3201,
NGC 4372, NGC 6362, NGC 6752, $\omega$~Centauri (NGC 5139) and 47
Tucanae (NGC 104). The discovery of two DNe, namely M55-CV1 and
M22-CV2, was already reported by \citet{kal05a} and \citet{pie05},
respectively. In the remaining 14 GCs we found no certain new DNe.
Our result raises the total number of known DNe in the Galactic
globular clusters to 12 DNe, distributed among 7 clusters. Our
survey recovered all three already known erupting cataclysmic
variables (CVs) located in our fields, namely M5-V101, M22-CV1,
and V4 in the foreground of M30. To assess the efficiency of the
survey, we analyzed images with inserted artificial stars mimicking
outbursts of the prototype dwarf novae SS~Cygni and U~Geminorum.
Depending on the conditions, we recovered between 16-100\% of
these artificial stars. The efficiency seems to be predominantly
affected by duty cycle/time sampling and much less by
distance/magnitude. Except for saturated tiny collapsed cores of
M30, NGC~362 and NGC~6752 (and also the dense core of NGC~2808)
crowding effects in the $V$ band were avoided by our image subtraction
technique augmented with auxiliary unsaturated $B$-band images.
Our results clearly demonstrate that in GCs common types
of DNe are very rare indeed. However, great care must be taken
before these conclusions can be extended to the CV population
in GCs.
\end{abstract}



\section{Introduction}

The dense central regions of globular clusters constitute rich
environments for the study of dynamical processes in stellar
systems. It is well established on theoretical grounds that close
binary stars play a significant role in the dynamical evolution of
GCs \citep{stod86,hut92}. For example, the presence
of the binary stars can delay or halt the evolution of a cluster
toward high stellar density, which is known as 'core collapse'. The
gravitational encounters between passing binaries and single stars
may produce tightly bound systems forming a variety of exotic
objects. These include active and quiescent low-mass X-ray binaries
(LMXBs) with a neutron star primary, millisecond pulsars (MSPs),
cataclysmic variables (CVs) and blue stragglers (BSs).

Cataclysmic variables are interacting binaries containing a
main-sequence or slightly evolved secondary star losing mass via
Roche lobe overflow onto a white dwarf primary. The binaries with
the magnetic field of the white dwarf strong enough
($B>10^7$~G) to channel the mass-flow along the field lines
directly onto the surface of the white dwarf are called
AM~Her-type or polar CVs. In the non-magnetic systems
($B<10^5$~G), an accretion disc forms around the primary. In the
mildly magnetic systems, $10^5<B<10^7$~G, termed intermediate
polars or DQ~Her-type systems, the accretion disc is truncated by
the white dwarf's magnetic field at a distance from the white
dwarf. It is believed that the accretion disc thermal instability
is the cause of repetitive outbursts observed in some CVs called
dwarf novae (DNe). The electronic edition of The Catalog and Atlas
of Cataclysmic Variables \citep{dow01} contains 480 certain DNe
among 1117 reliable CVs. There are 86 and
50 certain polars and intermediate polars in the database,
respectively. Thus, one can say that almost half of all known
field CVs exhibit outbursts, whereas at least 1/8 is of a magnetic
nature. However, one should also stress that the sample of known
field CVs suffers from strong, heterogeneous and rather
unquantifiable selection effects \citep[see,
e.g.,][]{gaen04,pret07}.

The issue of presence and formation of CVs in globular clusters
has been extensively studied on theoretical grounds. According
to \citet{bai90}, we should not expect dynamically
formed CVs in clusters due to unstable mass transfer in such systems.
Later, \citet{dis94} predicted the existence
of more than 100 CVs in both 47~Tuc and $\omega$ Centauri
and several thousand in the Galactic globular cluster system.
Recently, \citet{iva06} investigated CV formation channels
in GCs in detail. Using numerical simulations they estimate
the total number of CVs in the core of 47~Tuc at the level of
$\approx200$ binaries.

The number of known CVs in GCs is small. Over the last few years
observations collected with the orbiting X-ray telescopes yielded
merely several tens of candidate CVs for Chandra
\citep[e.g.][]{hann05,hei05,lugg07} and XMM-Newton
\citep[e.g.][]{webb04,webb06}. At present, there are just a few
spectroscopically confirmed CVs in GCs: the dwarf nova V101 in M5
\citep{mar81}, CV1-4 in NGC 6379 \citep{grin95,edm99}, a CV in NGC
6624 \citep{deu99}, and the objects V1, V2 and AKO9 in 47~Tucanae
\citep{kni04}. Dwarf nova outbursts were observed only in 12
objects (see Table~1). Among the latter, two DNe, namely CV1 in
the globular cluster M55 and CV2 in M22, were discovered in the
extensive photometric survey by the CASE collaboration
\citep{kal05a,pie05}.

CVs constitute a heterogeneous class
of stars studied with an assortment of tools,
from X-ray and UV satellites, QSO/blue excess photometric and
emission line surveys and by variability surveys, but these
methods suffer from severe and varying  selection effects. The
magnitude of these effects has never been reliably
estimated, as illustrated by the astronomer's failure to either
prove or reject the hibernation hypothesis, postulating over
factor 100 increase of the space density of CVs \citep[e.g.][]{sha86}.
For these reasons the results of our search of dwarf nova outbursts
in the globular clusters should not be compared to studies of other CV
sub-types and/or by other methods. Additionally, certain subtypes of
DNe were identified only after a half century of efforts at many
observatories. Thus, despite the lack of evidence to show an
abundance of possibly numerous but rarely detected subtypes,
this scenario cannot be ruled out. Among others, this does apply
to WZ~Sge stars having outbursts separated by decades of years.

It must be stressed that a restricted survey still
suffices to claim that the properties of the field and cluster DNe
differ, provided certain conditions are met. Namely, it must
be proven that (i) the survey is sensitive to detection of a
popular group of DNe, (ii) their relative counts in two samples
differ markedly. Thus it is our aim to demonstrate that our survey
is sensitive to SS~Cyg/U~Gem-type stars and that their count 
observed by us in the globular clusters is below expectations.
While our survey is restricted in the above sense, it must be
stressed that it is based on the most extensive photometric survey
of the globular clusters yet available. 

\begin{table*}
\begin{center}
\caption{All known DNe in Galactic globular clusters,
listed according to chronology of their discoveries.
Distances from cluster centres are given in four different
units: arcminutes, core radius $r_c$, half-mass radius $r_h$,
and tidal radius $r_t$.}
{\small
\begin{tabular}{|l|c|c|c|c|c|c|l|}
\hline
Dwarf Nova     &  RA(2000.0)  & Dec(2000.0) & \multicolumn{4}{|c|}{Distance from cluster centre}
                                                                                  & References  \\
               &   [h:m:s]    & [$^{\circ}:':''$] & ['] & [$r_c$] & [$r_h$] & [$r_t$] & \\
\hline
M5-V101        & 15:18:14.46  & +02:05:35.2 & 4.87 & 11.6~~ & 2.3~~~~ &  0.17~~~~ &  \citet{oost41} \\
47Tuc-V2       & 00:24:06.02  & -72:04:55.8 & 0.22 & ~0.55  & 0.08~~  &  0.005~~  &  \citet{par94} \\
M15-CV1(A)     & 21:29:58.32  & +12:10:01.3 & 0.01 & ~0.10  & 0.007   &  0.0005   &  \citet{cha02} \\
M22-CV1        & 18:36:24.66  & -23:54:35.5 & 0.40 & ~0.28  & 0.12~~  &  0.014~~  &  \citet{and03} \\
47Tuc-AKO9     & 00:24:04.95  & -72:04:55.2 & 0.09 & ~0.23  & 0.03~~  &  0.0022   &  \citet{kni03} \\
M55-CV1        & 19:40:08.59  & -30:58:51.1 & 2.56 & ~0.90  & 0.88~~  &  0.16~~~~ &  \citet{kal05a} \\
NGC6397-CV2    & 17:40:42.21  & -53:40:28.7 & 0.24 & ~4.71  & 0.10~~  &  0.015~~  &  \citet{sha05}, \citet{kal06} \\
NGC6397-CV3    & 17:40:42.48  & -53:40:17.4 & 0.32 & ~6.42  & 0.14~~  &  0.020~~  &  \citet{sha05} \\
M80-DN1        & 16:17:02.2~~ & -22:58:37.9 & 0.15 & ~1.01  & 0.23~~  &  0.011~~  &  \citet{shz05} \\
M80-DN2        & 16:16:59.8~~ & -22:58:18.0 & 0.70 & ~4.69  & 1.08~~  &  0.053~~  &  \citet{shz05} \\
M15-C          & 21:29:57.34  & +12:10:43.7 & 0.76 & 10.8~~ & 0.72~~  &  0.035~~  &  \citet{hann05} \\
M22-CV2        & 18:36:02.72  & -23:55:24.6 & 5.51 & ~3.88  & 1.69~~  &  0.19~~~~ &  \citet{pie05} \\
\hline
\end{tabular}}
\end{center}
\end{table*}

In this paper we present the results of our search for DNe by
CASE in the 16 Galactic globular clusters. The observation
and reduction procedures are described in Sect. 2. The
data and results for the individual clusters are presented in
Sect. 3. Simulations were performed to assess the completeness of
our search. We inserted artificial DNe into the frames of
three selected clusters. Results of these simulations
are presented in Sect. 4. In Sect. 5 we discuss implications of
our search for DNe in GCs, and our summary and conclusions are
presented in Sect. 6.


\section{Observations and data analysis}

The CASE project commenced at Las Campanas Observatory in 1996 and is
continuing until now \citep{kal05b}. For the survey we employ the
1.0-m Swope telescope. In 1996 only the globular cluster NGC 6752
was observed using two CCD cameras: FORD and SITE1. All observations
since 1997 have been obtained with a $2048 \times 3150$ pixel SITE3
CCD camera. With a scale of 0.435 arcsec/pixel the field of view
is $14\farcm8 \times 23\arcmin$. However, images of clusters were
usually taken with a smaller sub-raster (see Table 3 for details).
The observations of the globular cluster 47~Tucanae (NGC 104) were
obtained in the year 1993 as a by-product of the early stage of
Optical Gravitational Lensing Experiment \citep[OGLE,][]{uda92}
using the Swope telescope. Some GCs were also observed with the
TEK5 direct CCD camera attached to the 2.5-m du~Pont telescope.

Over 25,000 frames in total were obtained for the CASE survey.
Most of them, usually between 75 and 95\% per cluster, were taken
in the Johnson $V$ filter. The rest of the frames were taken through
the Johnson $B$ filter and rarely through Johnson $U$ and Cousins $I$
filters. The exposure times ranged from a few seconds to 600 s. For
the present search of DNe we used only $V$-band images with the
exposure times in excess of 80 s, numbering $\sim$19,700. The
number of $V$-band frames taken per cluster per night ranged from
1 to 116. The best seeing measured in this band reached 0.97
arcsec. 

Table~2 lists basic observational and physical parameters of the
16 observed globular clusters. In Table~3 we give general information
on analysed data. Note that for 47~Tucanae and $\omega$~Centauri,
two separate fields were monitored. Effectively we have no $V$-band
data on the very small cores of collapsed clusters M30, NGC 362 and
NGC 6752. Their cores of radii listed in Table~2 were saturated on
most of our deep frames (Sect.~3.6, 3.9 \& 3.14). This deficiency is
largely compensated for NGC 362 (and also NGC 2808), whose central
field $2 \farcm 9\times2 \farcm 9$ was observed in the $B$ band.
In the $B$ band giants in the collapsed core do not saturate and
blue outbursts of DNe would be easily detected.

\begin{table*}
\begin{center}
\caption{General information on analysed globular clusters. All values
without footnotes, except distances $d$ (column 6) and expected number
of CVs in cores (column 9), were taken from \citet{harr96}.
The distances were calculated basing on distance modulus $(m-M)_V$ and
reddening $E(B-V)$, and assuming absorption $A_V=3.2 \cdot E(B-V)$.
The predicted number of CVs in cores of the clusters was scaled
to $N_{CV}=200$ for 47~Tucanae according to formula,
given by \citet{pool03}, that the encounter rate
$\Gamma \propto \rho _0^{1.5} r_c^2$, where $\rho _0$ is
the central luminosity density \citep[also adopted from][]{harr96}.}
\begin{tabular}{|r|c|c|c|c|c|c|c|c|c|}
\hline
   NGC & Common       &  $V$  & $E(B-V)$  &      $(m-M)_V$       &      $d$      &       [Fe/H]       &   $r_c$   &   Expected number  & Remarks \\
       & name         & [mag] &   [mag]   &        [mag]         &     [kpc]     &                    & [\arcmin] & of CVs in the core & \\
\hline
   104 & 47 Tuc       & 3.95 &  0.02$^a$  &  $13.14 \pm 0.25^b$  & $4.1 \pm 0.5$ & $-0.67 \pm 0.05^c$ & 0.40 &   200        & \\
   288 &              & 8.09 &  0.03~     &  $14.57 \pm 0.07^d$  & $7.8 \pm 0.3$ & $-1.39 \pm 0.01^e$ & 1.42 &   ~~~~~~~0.3 & \\
   362 &              & 6.40 &  0.05~     &  $14.50 \pm 0.07^f$  & $7.4 \pm 0.2$ & $-1.33 \pm 0.01^e$ & 0.19 &   101        & ccc(?) \\
  2808 &              & 6.20 &  0.22~     &  $15.56 \pm 0.14^g$  & $9.4 \pm 0.6$ & $-1.36 \pm 0.05^h$ & 0.26 &   208        & \\
  3201 &              & 6.75 &  0.28$^i$  &  $14.08 \pm 0.16^j$  & $4.7 \pm 0.4$ & $-1.42 \pm 0.03^k$ & 1.43 &   ~~~~~~~2.6 & \\
  4372 &              & 7.24 &  0.34$^l$  &  $15.01$             & $6.1$         & $-2.10 \pm 0.05^m$ & 1.75 &   ~~~~~~~0.7 & \\
  5139 & $\omega$ Cen & 3.68 &  0.12~     &  $14.09 \pm 0.04^n$  & $5.5 \pm 0.1$ & $-1.7 \div -0.6^o$ & 1.40 &   ~~30       & \\
  5904 & M 5          & 5.65 &  0.03~     &  $14.56 \pm 0.10^p$  & $7.8 \pm 0.4$ & $-1.11 \pm 0.03^q$ & 0.42 &   ~~36       & \\
  6121 & M 4          & 5.63 &  0.33$^r$  &  $12.51 \pm 0.09^s$  & $1.73^s$      & $-1.17 \pm 0.31^t$ & 0.83 &   ~~~~~~~4.8 & \\
  6218 & M 12         & 6.70 &  0.19~     &  $14.22 \pm 0.11^u$  & $5.3 \pm 0.3$ & $-1.40 \pm 0.07^v$ & 0.72 &   ~~~~~~~4.6 & \\
  6254 & M 10         & 6.60 &  0.28~     &  $14.23 \pm 0.11^g$  & $4.6 \pm 0.2$ & $-1.52 \pm 0.02^w$ & 0.86 &   ~~13       & \\
  6362 &              & 7.73 &  0.09~     &  $14.46 \pm 0.10^x$  & $6.8 \pm 0.3$ & $-1.18 \pm 0.06^v$ & 1.32 &   ~~~~~~~0.9 & \\
  6656 & M 22         & 5.10 &  0.38$^y$  &  $13.74 \pm 0.20^z$  & $3.4 \pm 0.3$ & $-1.68 \pm 0.15^z$ & 1.42 &   ~~30       & \\
  6752 &              & 5.40 &  0.05$^*$  &  $13.24 \pm 0.08^a$  & $4.1 \pm 0.2$ & $-1.48 \pm 0.07^*$ & 0.17 &   ~~51       & ccc \\
  6809 & M 55         & 6.32 &  0.08~     &  $13.86 \pm 0.25^\#$ & $5.3 \pm 0.6$ & $-1.71 \pm 0.04^y$ & 2.83 &   ~~~~~~~1.7 & \\
  7099 & M 30         & 7.19 &  0.03~     &  $14.65 \pm 0.12^\&$ & $8.1 \pm 0.5$ & $-2.01 \pm 0.09^\&$& 0.06 &   ~~39       & ccc \\
\hline
\end{tabular}
\end{center}
\vspace*{-0.3cm}
{\footnotesize References: $^a$~\citet{gra03}, $^b$~\citet{wel04}, $^c$~\citet{carr04},
$^d$~\citet{chen00}, $^e$~\citet{she00}, $^f$~\citet{ferr99}, $^g$~\citet{saad01},
$^h$~\citet{walk99}, $^i$~\citet{lay02}, $^j$~\citet{maz03}, $^k$~\citet{gon98},
$^l$~\citet{ger04}, $^m$~\citet{gei95}, $^n$~\citet{kal02}, $^o$~\citet{nor04},
$^p$~\citet{lay05}, $^q$~\citet{carr97}, $^r$~\citet{ivan99}, $^s$~\citet{rich97},
$^t$~\citet{dra94}, $^u$~\citet{har04}, $^v$~\citet{rut97}, $^w$~\citet{kra95},
$^x$~\citet{ole01}, $^y$~\citet{richt99}, $^z$~\citet{mon04},
$^*$~\citet{gra05}, $^\#$~\citet{pych01}, $^\&$~\citet{sand99}. \\
Remark: ccc - core-collapsed cluster.}
\end{table*}

\begin{table*}
\begin{center}
\caption{Summary of observational data on analysed globular clusters.
All clusters (except 47~Tuc) were observed with the SITE3 CCD camera.
All fields (except 47~Tuc A,B and $\omega$ Cen E,W) were centered on
the clusters.}
{\small
\begin{tabular}{|l|c|c|c|c|c|c|}
\hline
                 &                       & Total number &   Total    &       Size of      & Detection  & Limiting \\
Analysed         &      Seasons          &   of long    & number of  &      analysed      &   limit    & absolute \\
field            &                       &  exposures   &  analysed  &        field       &   in $V$   & magnitude \\
                 &                       & in $V$ band  &   nights   &    [arcmin$^2$]    &   [mag]    &  $M_V$  \\
\hline
M 4              & 1998-2000, 2002-2005  &    1981      &    ~~61    & $14.8 \times 14.8$ &   20.2     &     7.7 \\
M 5              & 1997-1999, 2002-2004  &    1043      &    ~~34    & $11.6 \times 14.8$ &   21.1     &     6.5 \\
M 10             &      1998, 2002       &   ~~847      &    ~~28    & $13.6 \times 14.8$ &   20.5     &     6.3 \\
M 12             &      1999-2001        &    1236      &    ~~41    & $14.8 \times 11.6$ &   20.5     &     6.3 \\
M 22             &      2000-2001        &    2006      &    ~~71    & $14.8 \times 11.6$ &   20.1     &     6.5 \\
M 30             &        2000           &   ~~340      &    ~~23    & $14.8 \times 14.8$ & ~~21.1$^*$ &   ~~6.5$^*$ \\
M 55             &      1997-2004        &    3795      &     151    & $14.8 \times 11.6$ &   20.5     &     6.6 \\
NGC 288          &      2004-2005        &   ~~297      &   ~~~~9    & $11.9 \times 11.9$ &   20.9     &     6.3 \\
NGC 362          & 1997-1998, 2000-2005  &    1424      &    ~~90    & $14.8 \times 11.6$ & ~~20.6$^*$ &   ~~6.1$^*$ \\
NGC 2808         &      1998-1999        &   ~~312      &    ~~33    & $14.8 \times 14.8$ & ~~20.5$^*$ &   ~~4.9$^*$ \\
NGC 3201         &      2001-2005        &   ~~751      &    ~~22    & $11.9 \times 14.9$ &   20.8     &     6.7 \\
NGC 4372         &      2004-2005        &   ~~601      &    ~~19    & $11.9 \times 14.9$ &   20.4     &     5.4 \\
NGC 6362         &      1999-2005        &    2585      &     104    & $14.8 \times 11.6$ &   20.4     &     5.9 \\
NGC 6752$^a$     &      1996-1997        &   ~~395      &   ~~~~7    & $14.8 \times 14.8$ & ~~20.5$^*$ &   ~~7.3$^*$ \\
$\omega$ Cen E   &      1999-2001        &   ~~815      &    ~~68    & $14.8 \times 22.8$ &   20.2     &     6.1 \\
$\omega$ Cen W   &      1999-2002        &   ~~756      &    ~~76    & $14.8 \times 22.8$ &   20.4     &     6.3 \\
47 Tuc A$^b$     &        1993           &   ~~277      &    ~~34    & $14.8 \times 14.8$ &   21.5     &     8.4 \\
47 Tuc B$^b$     &        1993           &   ~~274      &    ~~30    & $14.8 \times 14.6$ &   21.5     &     8.4 \\
\hline
\end{tabular}}
\end{center}
\vspace*{-0.3cm}
{\footnotesize Remarks: $^a$~cluster observed with the SITE1, SITE3 and FORD
CCD cameras, and not analysed with the subtraction method, \\
$^b$~cluster observed only with the LORAL CCD camera, \\
$^*$~due to small and crowded core the magnitude applies only to area
outside the core}
\end{table*}

In our search procedure we used the Difference Image Analysis Package
(DIAPL). The package was written by \citet{woz00} and recently modified by
W. Pych \footnote {The package is available at\\
http://www.camk.edu.pl/$\sim$pych/DIAPL}. It is an implementation of
the method developed by \citet{ala98}. To get better quality
of photometry each frame was split into $410 \times 410$ pixel subfields.

For each cluster we selected the nights where at least one
$V$-band image with seeing better than $1\farcs61$
(3.7 pixels for the SITE3 camera) was available.
Table~3 gives the total number of nights as well as the total number
of exposures per cluster used in the analysis. For each cluster
a reference image was constructed by combining
5 to 23 individual frames taken during dark time on one
selected night. For each night in the data set we combined up
to 5 of the best images (with good seeing and relatively low background)
to form an average image. We thus obtained a sequence of frames,
each representing one night. The combined frames were remapped
to the reference image coordinate system and subtracted
from the convolved reference image using DIAPL.
The resultant frames were searched with DAOPHOT \citep{stet87}
for the presence of any stellar-like residuals. Any residuals
which appeared in two subsequent frames (including breaks
in observations up to 4 days) and were separated by no more than
0.25 pixels ($\approx0\farcs109$) were selected for further
examination. We omitted from the search regions corresponding to the
locations of saturated stars or to known variables (more extended
lists based on CASE results will be published elsewhere).
Subsequently, we extracted light curves in units of residual counts
on subtracted images. This was done with DIAPL.
Finally, the light curves were examined by eye.

Such a procedure would fail to detect a DN in eruption on the reference
night due to negative residuals on subtracted images. For this reason
for each cluster we selected a subtracted image separated by 20-40 days
from the reference night. The image was inverted and then searched
for the presence of positive stellar-like residuals. For all positive
residuals we extracted light curves and examined them visually.


\section{The individual clusters}

\subsection{M4}

M4 is the closest globular cluster \citep[1.73 kpc,][]{rich97}.
The CASE observations of this cluster span the years 1998-2000 and
2002-2005. The total number of analysed nights reached 61,
including 25 of the year  1998. The distribution of the data in time is
presented in Fig.~1. Our search for DNe in this cluster yielded no
detections. Additionally, in subtracted images we inspected
visually the locations of two X-ray sources classified by
\citet{bas04} as possible CVs, namely CX1 and CX4.
No variability was found within $3\sigma$ error circles
around the sources.

\begin{figure}
\vspace{7.3cm}
\includegraphics{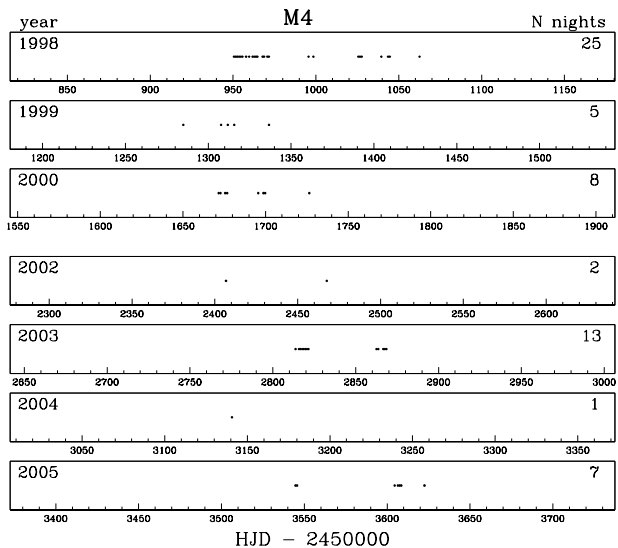}
\caption{\small Analysed observations of the globular cluster M4.}
\label{fig1}
\end{figure}

\subsection{M5}

M5 is the only northern cluster observed by the CASE team.
The observations cover 34 nights in the years 1997-1999 and 2002-2004
(Fig.~2). Our outburst search procedure easily detected
the known DN M5-V101. In Fig.~3 we present its light curve.
Apart from two eruptions in 1997, reported by \citet{kal99},
we detected two more outbursts in 2003 and in 2004.
No new erupting objects were found in the field of M5.

\begin{figure}
\vspace{6.3cm}
\includegraphics{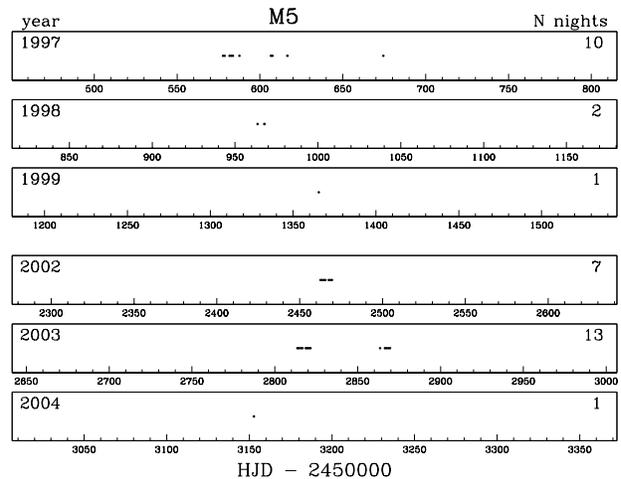}
\caption{\small Graphical log of analysed observations of the globular
cluster M5.}
\label{fig2}
\end{figure}

\begin{figure}
\vspace{2.0cm}
\includegraphics{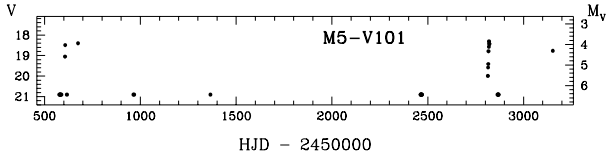}
\caption{\small Light curve of the DN V101 in the globular cluster M5.
The curve was extracted using the DIAPL package.
Each point represents one night.}
\label{fig3}
\end{figure}

\subsection{M10}

This cluster was monitored in 1998 (18 nights) and in 2002 (10
nights). Distribution of observations in time is presented in Fig.~4.
No outbursting star was detected here.

\begin{figure}
\vspace{2.8cm}
\includegraphics{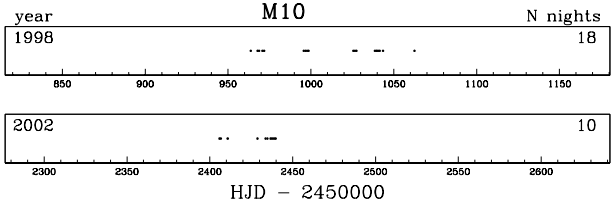}
\caption{\small Log of analysed observations of the globular cluster M10.}
\label{fig4}
\end{figure}

\subsection{M12}

M12, another globular cluster, was observed in the years 1999-2001
on 41 nights (Fig.~5). These observations were also searched for
DNe but unfortunately yielded no detection in the whole cluster.

\begin{figure}
\vspace{3.4cm}
\includegraphics{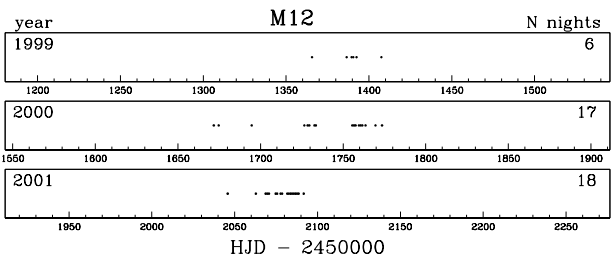}
\caption{\small Distribution of analysed observations of the globular
cluster M12.}
\label{fig5}
\end{figure}

\subsection{M22}

M22 was monitored during the 2000 and 2001 seasons (Fig.~6). The
results of a search for erupting objects in the field of this
cluster were published by \citet{pie05}, in which we reported
the discovery of the DN M22-CV2 (located at a distance of 3.9 core
radii from the centre of the cluster). The observations revealed an
SU~UMa type superoutburst in the year 2000. Prominent superhumps with
a period of 128 minutes were observed in the superoutburst light
curve during three nights. M22-CV2 has an X-ray counterpart
\citep[source \#40 in][]{webb04}, but the cluster membership
of the object remains an open question.
We also registered two outbursts of the previously known DN M22-CV1.
In Fig.~7 we show $V$-band light curves of the two CVs observed
in the field of M22.

\begin{figure}
\vspace{2.6cm}
\includegraphics{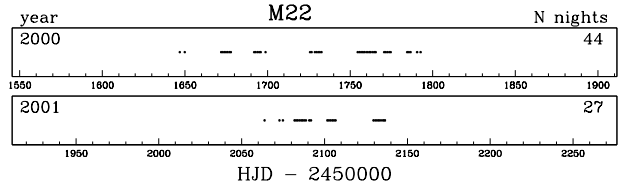}
\caption{\small Time-domain distribution of analysed observations of
the globular cluster M22.}
\label{fig6}
\end{figure}

\begin{figure}
\vspace{4.3cm}
\includegraphics{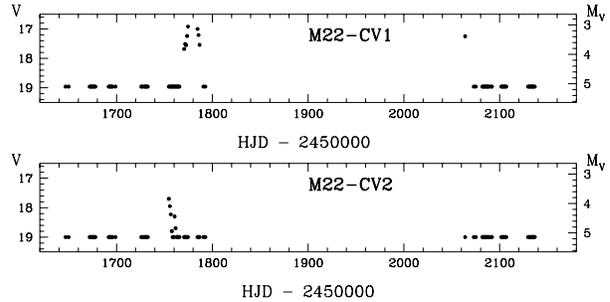}
\caption{\small Light curves of the DNe CV1 and CV2 in the
globular cluster M22 extracted with DIAPL package. Each point represents
one night. Note that for none of the CVs the actual brightness
maxima were observed.}
\label{fig7}
\end{figure}

\subsection{M30}

M30 is one of the 30 Galactic globular clusters which, according to
\citet{harr96}, has likely undergone core collapse. The CASE
observations of this cluster cover 23 nights between June and
September 2000 (Fig.~8).  The core of M30, extending by
$0\farcm06 \approx 8.3$ pixels and covering roughly 30 seeing 
disks, was mostly saturated on our deep frames. Because of effective 
lack of data there it remained ignored in our search of outbursts 
and simulations. 

Our search of DNe easily recovered the known DN
of U~Gem type labeled M30-V4 ($\alpha_{2000}=21^{h} 39^{m}
58.5^{s}$, $\delta_{2000} =-23\degr 11\arcmin 44\arcsec$). The
variable was discovered by \citet{ros49}. Later \citet{mach91}
found that, on the basis of its quiescent magnitude and
colours, its outburst magnitude and its radial velocity, V4 was
likely to be a foreground object in the field of M30. A relatively
high $V$-band brightness (see Fig.~9)
is consistent with its foreground location. It is worth
noting that M30-V4 has a likely X-ray counterpart, J2139.9-2312,
detected by the $ROSAT$ satellite \citep{whi00}.
The equatorial coordinates of the X-ray source are
$\alpha_{2000}=21^{h} 39^{m} 57.9^{s}$, $\delta_{2000} =-23\degr
12\arcmin 06\arcsec$, offset from the optical position of M30-V4
by ($\Delta \alpha=9\arcsec$, $\Delta \delta=22\arcsec$), well
within the $50\arcsec$ X-ray error radius.

Additionally, in subtracted images we inspected the locations of four
X-ray sources, namely A2, A3, B and C, which according to \citet{lugg07}
have X-ray properties consistent with being CVs. We also checked
two other bright X-ray sources, A1 (a very likely qLMXB)
and D (probably a quasar). No variability was found within
$3\sigma$ error circles around all these sources.

\begin{figure}
\vspace{1.6cm}
\includegraphics{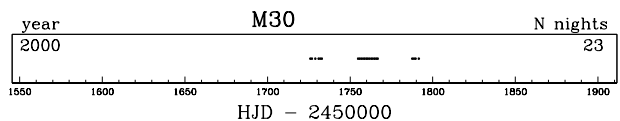}
\caption{\small Analysed observations of the globular cluster M30.}
\label{fig8}
\end{figure}

\begin{figure}
\vspace{2.0cm}
\includegraphics{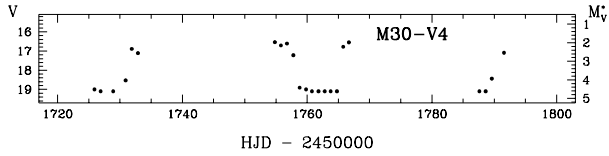}
\caption{\small Light curve of the DN V4 in the field of the
globular cluster M30. Each point represents one night. At the cluster
distance these values would correspond to the absolute magnitudes 
$M_V^*$ given on the right vertical axis.
They would be excessively bright for a dwarf nova, hence
it is extremely likely this is a foreground dwarf nova.}
\label{fig9}
\end{figure}

\subsection{M55}

The time distribution of the observations in Fig.~10 reveal that
M55 was observed most often in the whole CASE survey. The results of
our search for DNe outbursts were presented by \citet{kal05a}.
In that paper we reported the discovery of
the DN M55-CV1 in the core of the cluster. Over eight observing
seasons spanning the period 1997-2004 we observed six outbursts of
this variable star. The X-ray flux as well as the star location in
the color-magnitude diagram for M55 are consistent with it being a
DN at the cluster distance. In Fig.~11 we present the light curve
of M55-CV1 in magnitudes.

\begin{figure}
\vspace{8.1cm}
\includegraphics{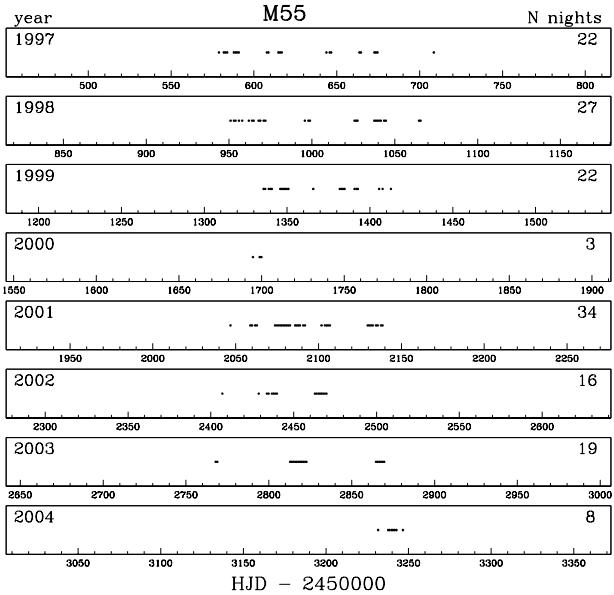}
\caption{\small Distribution of the analysed observations of the globular
cluster M55.}
\label{fig10}
\end{figure}

\begin{figure}
\vspace{2.0cm}
\includegraphics{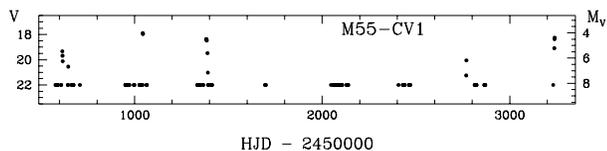}
\caption{\small Light curve of the DN M55-CV1.
Each point represents one night.}
\label{fig11}
\end{figure}

\subsection{NGC 288}

The globular cluster NGC 288 is located about 1 deg from the
Southern Galactic Pole. Among all analysed clusters it is the most
distant from the Galactic Centre ($\sim$12 kpc). The CASE
observations cover only two nights in 2004 and 7 nights in 2005
(Fig.~12). Our search for erupting objects yielded a negative result.
We also carried out an independent visual inspection of the
regions of optical counterparts to seven X-ray sources,
CX13, CX15, CX18, CX19, CX20, CX24, CX25, listed by \citet{kong06}.
Three of the sources, namely CX13, CX20 and CX24, are CV candidates.
No brightness variations were found at the positions of all these objects.

\begin{figure}
\vspace{2.6cm}
\includegraphics{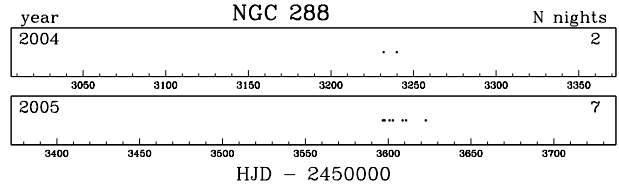}
\caption{\small Graphical log of analysed observations of NGC 288.}
\label{fig12}
\end{figure}

\subsection{NGC 362}

The globular cluster NGC 362 lies on the sky about 2 deg away from
the centre of Small Magellanic Cloud (SMC). It is a likely
core-collapsed cluster. The CASE data for NGC 362 were obtained
from 1997 until 2005, except for the 1999 season. The total number
of analysed nights reached 90 (Fig.~13). Our search of CV outbursts
yielded no detections. In the $V$ band the collapsed core of radius 
$11\farcs4$ and roughly covering 300 seeing radii was saturated, 
yielding no useful data. This was largely compensated by our analysis
of the $B$-band frames of the central field of the cluster,
covering $2 \farcm 9\times2 \farcm 9$. In this filter red giants do not
saturate in the core and any erupting CVs should become more apparent
due to their hot and luminous disc in the systems. The result of
our search for this cluster is again negative.

\begin{figure}
\vspace{8.1cm}
\includegraphics{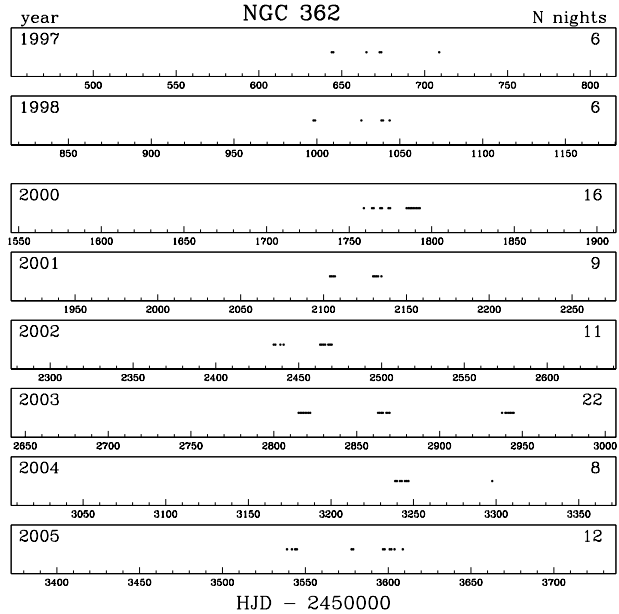}
\caption{\small Time-domain distribution of analysed observations
of the globular cluster NGC 362.}
\label{fig13}
\end{figure}

\subsection{NGC 2808}

NGC 2808 is the most distant globular cluster in our sample
\citep[$9.4\pm0.6$ kpc, $(m-M)_V=15.56$ mag,][]{saad01}. The
cluster was monitored in 1998 and 1999 during 33 nights in total
(Fig.~14). For this cluster we analysed the large $14 \farcm
8\times14 \farcm 8$ field in the $V$ band and the $2 \farcm 9\times2
\farcm 9$ central field in the $B$ band. The results of our search
yielded no DN outbursts for this cluster, too.

\begin{figure}
\vspace{2.6cm}
\includegraphics{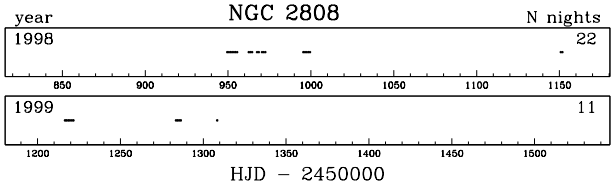}
\caption{\small Analysed observations of the globular cluster NGC 2808.}
\label{fig14}
\end{figure}

\subsection{NGC 3201}

In our sample of 16 GCs, NGC 3201 is one of three clusters located
closer than 10 deg from the Galactic plane. The estimated
value of the reddening in the direction of the cluster amounts to
$E(B-V)=0.23$ \citep{lay02}. Although our observations of NGC 3201
span five years, 2001-2005, they cover only 22 nights (Fig.~15).
We found no erupting objects in this cluster either.
Additional visual inspection of the locations of four X-ray
sources, 16, 22, 23, 26 (which according to \citet{webb06}
could be CVs) revealed no brightness variations.
We also note no variability within $3\sigma$ uncertainty
circles around other 23 X-ray sources located in our field
of view for NGC 3201.

\begin{figure}
\vspace{4.9cm}
\includegraphics{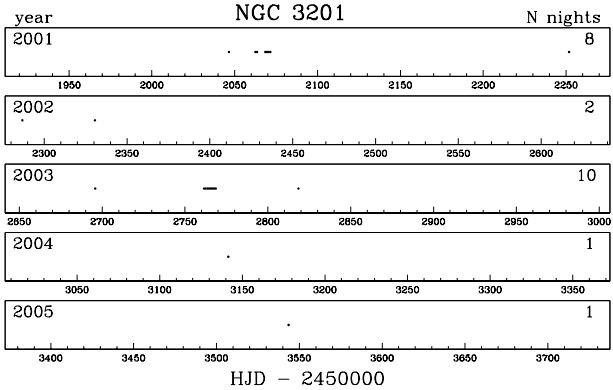}
\caption{\small Analysed observations of NGC 3201.}
\label{fig15}
\end{figure}

\subsection{NGC 4372}

This globular cluster is located relatively close to the Galactic
plane ($b=-9 \fdg 88$) and suffers from reddening of $E(B-V)=0.34$
\citep{ger04}. Images of NGC 4372 were obtained on 14
nights in 2004 and on 5 nights in 2005 (Fig.~16). No DNe were detected in
this cluster.

\begin{figure}
\vspace{2.6cm}
\includegraphics{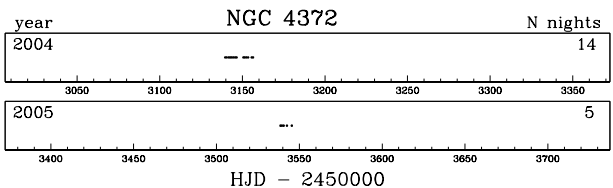}
\caption{\small Distribution of analysed data of the globular cluster NGC4372.}
\label{fig16}
\end{figure}

\subsection{NGC 6362}

The analysed observations of NGC 6362 span six years, from 1999 to
2005. The total number of nights reached 104, two thirds of which
dates to 1999 and 2001 seasons (Fig.~17). Despite the extensive
observations of this cluster no DN outburst was detected.

\begin{figure}
\vspace{6.9cm}
\includegraphics{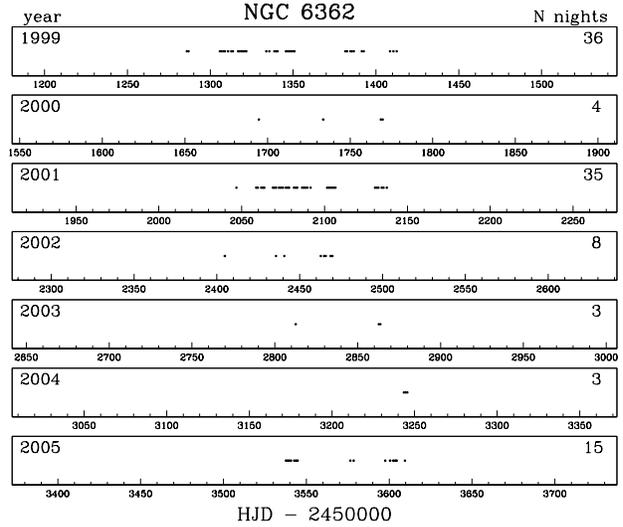}
\caption{\small Graphical log of observations for the globular cluster NGC6362.}
\label{fig17}
\end{figure}

\subsection{NGC 6752}

NGC 6752 was observed using three different CCD cameras (FORD,
SITE1 and SITE3) on only 7 nights in 1996 and 1997 (see Fig.~18).
Because of the difference in our frame format, for this cluster
we performed only visual examination. For each night we selected
the best seeing $B$ and $V$ images and subsequently we examined
them by eye. None of the 11 optical identifications of X-ray
sources (CX1, CX2, CX3, CX4, CX5,
CX6, CX7, CX10, CX11, CX13, CX15) listed by \citet{pool02} as
likely CVs, revealed any brightness variations within 3$\sigma$
error circles. Note that in the $V$ band the collapsed core of
radius $10\farcs2$ and covering less than 300 seeing disks
was saturated and yielded no useful data.

\begin{figure}
\vspace{2.6cm}
\includegraphics{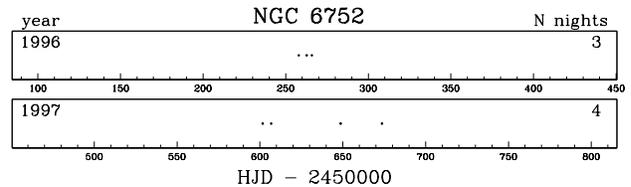}
\caption{\small Time-domain distribution of analysed observations of
the globular cluster NGC6752.}
\label{fig18}
\end{figure}

\subsection{$\omega$ Centauri}

$\omega$ Centauri is the most luminous and the most massive of all
known Galactic GCs. Two partly overlapping fields, East and West,
covering the central region of the cluster were monitored. The
full size of the surveyed area was equal to 644 arcmin$^2$. The
CCD photometry of the two fields was carried out from February
1999 until June 2001. Additionally, the West field was observed in
May and June 2002 (Fig.~19). Our extensive search of erupting
objects in $\omega$~Centauri led to the detection of a source
suffering from intriguing episodes of enhanced luminosity.
The source was located in the West field at $\alpha_{2000}=13^{h}
26^{m} 26\fs61$, $\delta_{2000}=-47\degr 26\arcmin 35\farcs1$.
Fig.~21 presents the finding chart centered on its location.
The object was already recorded in the catalogue of variable stars
published by \citet{kal04}. They suggest that the object
(labelled as the variable NV408) is a probable CV. The object
appeared on images obtained since June 3/4, 1999. It faded
systematically during six consecutive nights. On July 24/25, 1999
it was bright again, albeit not as much as the maximum in June.

\begin{figure}
\vspace{5.9cm}
\includegraphics{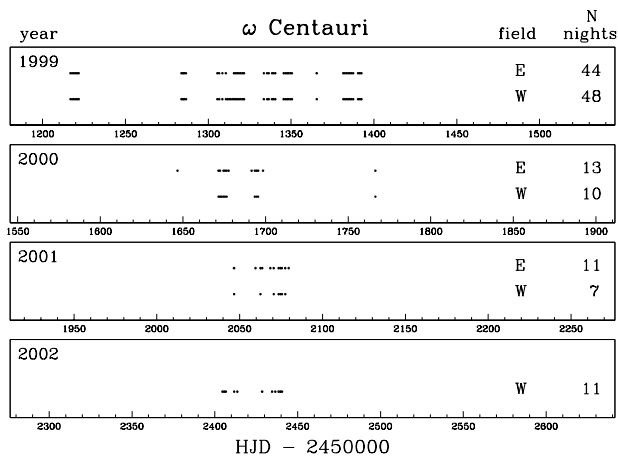}
\caption{\small Distribution of the analysed observations of the globular
cluster $\omega$~Centauri.}
\label{fig19}
\end{figure}

\begin{figure}
\vspace{2.0cm}
\includegraphics{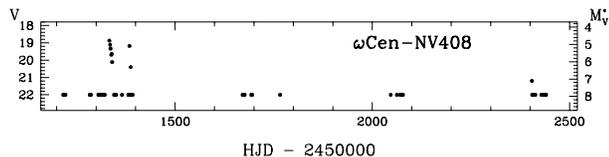}
\caption{\small Light curve of the variable object NV408
in the globular cluster $\omega$~Centauri. Each point represents one night.
If the object is the cluster member then it would have absolute magnitudes
$M_V^*$ given on the right vertical axis.}
\label{fig20}
\end{figure}

\begin{figure*}
\vspace{8.5cm}
\includegraphics{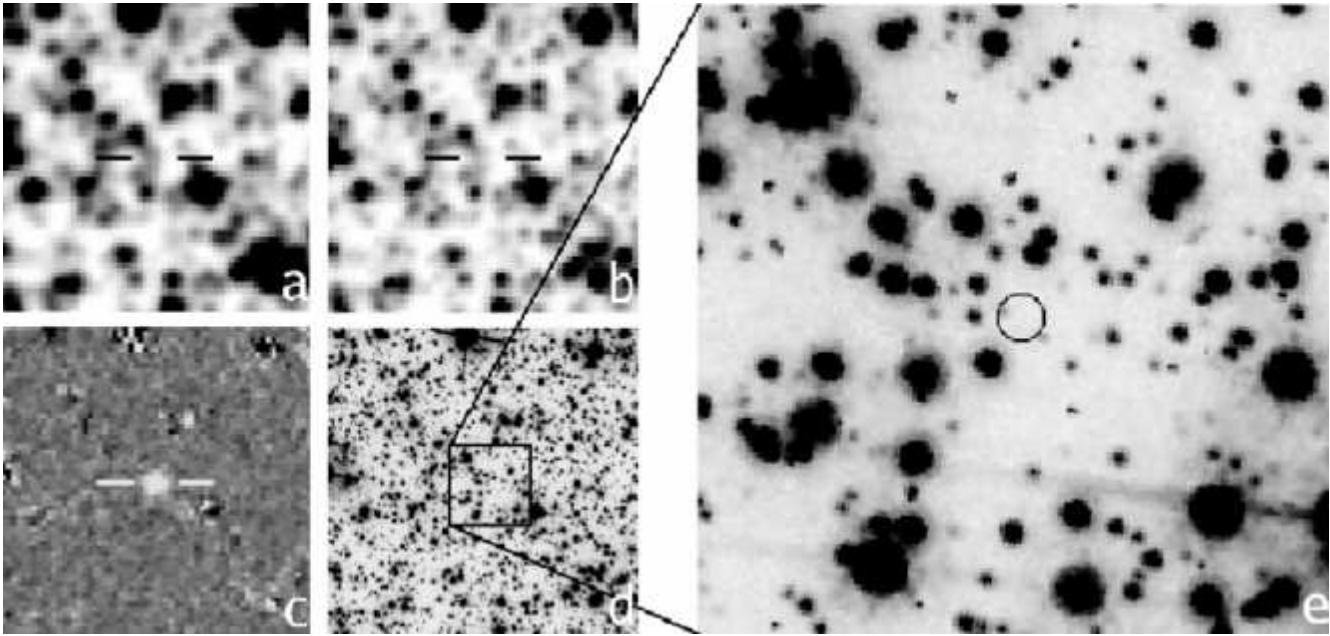}
\caption{\small Finding charts for the variable object NV408 in
$\omega$~Centauri. Each chart from 'a' to 'd' is $26\arcsec$ on a side.
North is up and East to the left. The charts 'a' and 'b' show the region
of NV408 during eruption (1999 June 3/4) and in quiet state
(1999 May 22/23), respectively. The chart 'c' shows the residual image
corresponding to the night of June 3/4, 1999. The HST/ACS image of the same
area, taken on June 28, 2002 in the F625W filter, is displayed in the
chart "d". The close-up view presented in panel "e" is $6\farcs7$
on a side. The 3-sigma circle with radius of $0\farcs26$ is centered
on the NV408 position. The faintest stars on the HST image have $R\sim26$ mag.
}
\label{fig21}
\end{figure*}

Our evidence is insufficient to prove that NV408 is an erupting
CV. It remains undetectable on our images taken with the 2.5-m du~Pont
telescope. In order to check for other signatures of a CV at this location,
we investigated the archival HST/ACS
images covering the region of NV408. These are images in F435W
(blue, in this subsection $B$) and F625W (red, $R$)
wide-band filters, and F658N narrow-band filter centered
on H$\alpha$ line, all taken on June 28, 2002. These checks
returned no object with a blue excess nor H$\alpha$ emission
within $2\arcsec$ radius from our derived position. Based on the paper
by \citet{mone05} we estimated the limiting magnitude for
the HST/ACS images to be $B\approx 27$ ($R\approx 26.2$).
At an apparent distance modulus to $\omega$~Centauri
$(m-M)_B\approx 14.2$ this yields the limiting absolute
magnitude $M_B\approx 12.8$.

We also note that NV408 has no X-ray counterpart.
No gamma ray burst was listed near its position in
The BATSE Current Gamma-Ray Burst Catalog
\footnote {The catalog can be found at\\
http://f64.nsstc.nasa.gov/batse/grb/catalog/current/} during its
brightening between May and July 1999. Thus, the nature of NV408
still remains unsolved.

We report no brightness variations in the vicinity of three
possible CV locations in $\omega$~Centauri. These are sources
NGC5139-A and NGC5139-B \citep{car00},
and NGC 5139-31 \citep{rut02}.

\subsection{47 Tucanae}

Similar to NGC 362, the globular cluster 47~Tucanae is also
projected onto the halo of the SMC. The observations of 47~Tucanae
(NGC 104) were obtained in the year 1993 as a
by-product of the Optical Gravitational Lensing Experiment
\citep[OGLE,][]{uda92}, at the stage when the 1.0-m Swope telescope
was used. We analysed fields A and B located west and east of the
cluster center, respectively, but the cluster core was not included
in these fields. All images of 47~Tuc were taken with
a $2048 \times 2048$ pixel LORAL CCD camera with a scale of 0.435
arcsec/pixel. The distribution of our observations in time is
presented in Fig.~22. Our search of DNe in this cluster yielded no
detections.

\begin{figure}
\vspace{2.0cm}
\includegraphics{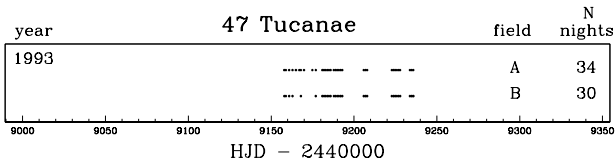}
\caption{\small Analysed observations of the globular cluster 47~Tucanae.}
\label{fig22}
\end{figure}


\section{Simulations}

Our extensive search for DNe in 16 Galactic GCs yielded only two new
objects, M22-CV2 \citep{kal05a} and M55-CV1 \citep{pie05} and recovered
three previously known DNe. While the recovery of known DNe lends some
confidence in the sensitivity of our search for outbursts, we performed
simulations to assess the completeness of our search in a more
quantitative way. We inserted into the frames of three clusters, M22,
M30, and NGC 2808, artificial images of erupting CVs and checked
whether they were detected in our search. The distance modulus
$(m-M)_V$ of the selected clusters increases from the lowest value
of $\sim13.7$~mag for M22 up to the highest value of $\sim15.6$~mag
for NGC 2808, the most distant cluster in our sample. The number of
observing nights per cluster was different for each of the three
clusters: 71 nights for two years for M22, 23 nights in one year
for M30, and 33 nights for two years for NGC 2808. Hence, M22
has the highest number of nights per year (35.5), while
NGC 2808 has the smallest (16.5).

In the simulations we reproduced the light curves of two prototype
DNe, SS~Cygni and U~Geminorum. The original light curves were taken
from the AAVSO International Database. Outbursts of SS~Cyg and U~Gem
repeat on average every 39 and 101 days, respectively \citep{war95}.
We selected the most recent and well-covered segments of the visual 
light curves of these DNe, illustrated in Fig.~23. 
The SS~Cyg light curve spans
from Aug 30, 2005 till Jan 30, 2006, while that for U~Gem from Sep 26,
2005 till May 30, 2006. Outbursts of SS~Cyg and U~Gem 
covered respectively $\sim$30\% and $\sim$10\% of their light curves. 
During the given time interval SS~Cyg
erupted six times, while U~Gem only twice.
These light curves were binned by calculating nightly averages.

\begin{figure}
\vspace{7.9cm}
\includegraphics{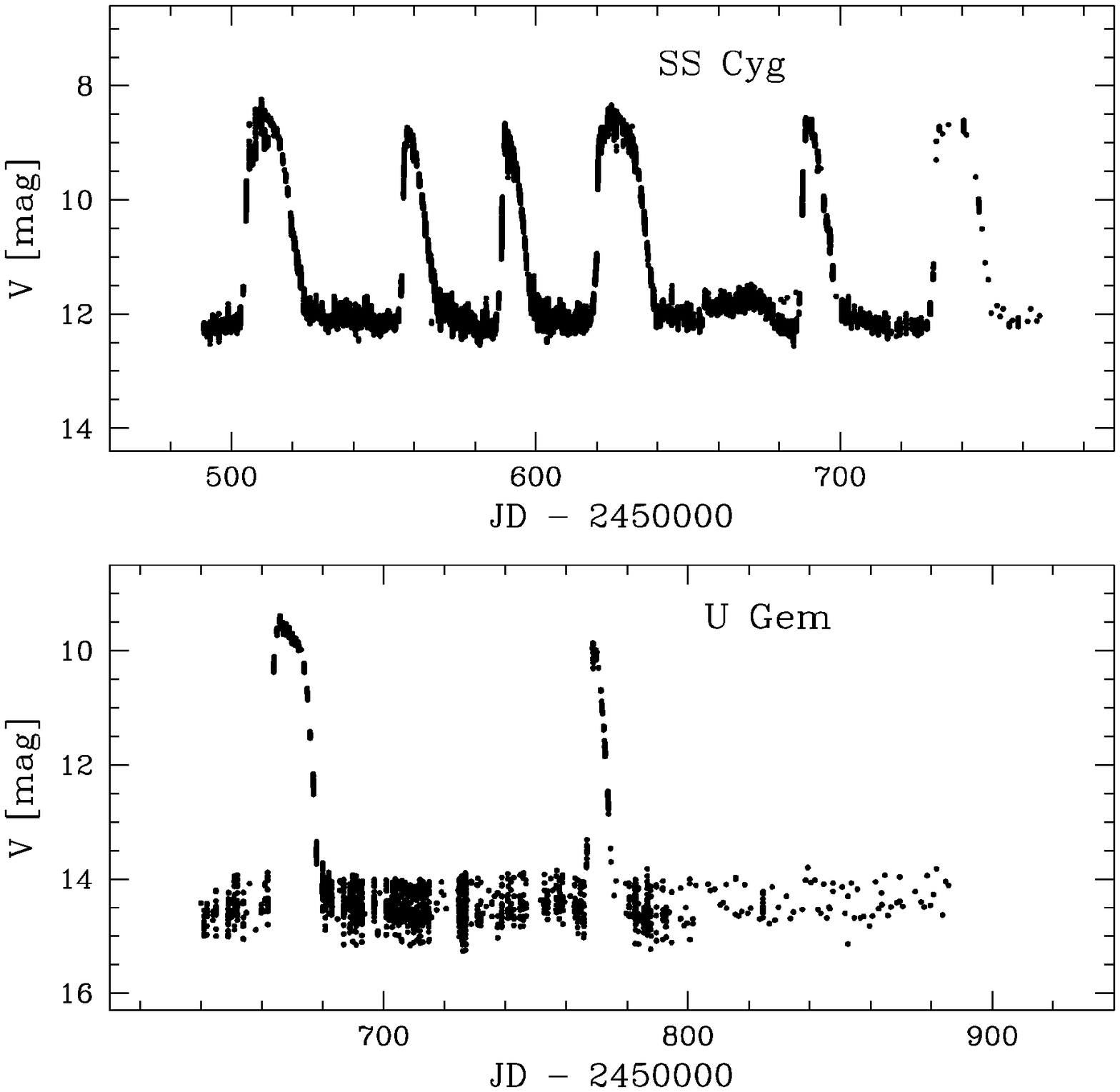}
\caption{\small AAVSO $V$-band light curves of the prototype DNe, SS~Cygni
and U~Geminorum, used in the simulations. The plots have the same scale
on both axes.}
\label{fig23}
\end{figure}

For each cluster the prototype light curves were scaled according
to its distance. In the process we used the recent
distance determinations of our prototype DNe by \citet{bas05}.
In Table~4 we list properties of these DNe and their hypothetical 
magnitude ranges in selected clusters.
We also added two other well-known DNe.
The absolute maximum brightness of SS~Cygni
exceeds by 2~mag that for the other three DNe. Only SS~Cyg
is sufficiently bright in its minimum to be detectable
at cluster distances in our images.

\begin{table*}
\begin{center}
\caption{Distances and $V$ magnitude ranges of four prototype DNe.
Absolute magnitudes are derived assuming zero absorption.
Last three columns give magnitudes of the prototype DNe
at the distance of the three globular clusters selected
for the simulations.}
{\small
\begin{tabular}{|c|c|c|c|c|c|c|}
\hline
Prototype & Apparent $V$ & Distance          &    $M_V$      & \multicolumn{3}{|c|}{Cluster} \\
    DN    &   [mag]      &  [pc]             &    [mag]      &    M22      &     M30     & NGC 2808 \\
\hline
U~Gem   & $~9.1~-~15.2$ & $96^{+5}_{-4}$     & $4.2~-~10.3$  & 17.9~-~24.0 & 18.9~-~25.0 & 19.8~-~25.9 \\
SS~Cyg  & $~8.2~-~12.1$ & $166.2\pm12.7$     & $2.1~-~~~6.0$ & 15.8~-~19.7 & 16.8~-~20.7 & 17.7~-~21.6 \\
WZ~Sge  & $~7.0~-~15.5$ & $43.5\pm0.3$       & $3.8~-~12.3$  & 17.5~-~26.0 & 18.5~-~27.0 & 19.4~-~27.9 \\
SU~UMa  & $11.2~-~15.0$ & $260^{+190}_{-90}$ & $4.1~-~~~7.9$ & 17.8~-~21.6 & 18.8~-~22.6 & 19.7~-~23.5 \\
\hline
\end{tabular}}
\end{center}
\end{table*}

Properties of our simulated artificial stars were randomised both
in time and space. For each simulated artificial star, we selected
a random time among the first 100 nights in the prototype light curve
and identified it with the actual time of first observation of a
given cluster in a given season. We simulated one point for each
night of the actual observations. Our simulations were performed in
subfields extending $410 \times 410$ pixel ($2\farcm97 \times
2\farcm97$). In each of them we inserted 10 SS~Cyg and 10 U~Gem
stars. Locations of these artificial stars were chosen randomly
within a given subfield. To add the artificial observation of a
DNe to a set of frames we employed the ADDSTAR task of DAOPHOT
\citep{stet87}. For this purpose we used the PSFs determined from
a given subfield, to ensure the correct shapes of the stellar images.

The complete set of simulated frames for each cluster was searched
for outbursts of DNe in exactly the same manner as the search in our
original data. This included additional visual inspection of
the light curves obtained as the result of image subtraction.
Example light curves of the detected artificial DNe are presented in
Fig.~24. Note that the candidate 'DNe' were easily identifiable due
to their characteristic pronounced flat minima with counts usually
near zero and outbursts reaching from hundreds to thousands of counts.

The results of the simulations are summarised in Table~5. They are
presented graphically in Fig.~25. In Table~5 we give the initial
number of synthetic DNe per cluster, the number (percentage) of
the DNe whose eruptions would be potentially detectable, and the 
number (percentage) of finally recovered DNe.  By 'potentially
detectable' we mean those which erupted during our observation. 
The difference between the number of DNe used in
the simulations for different clusters is the result of different
field of view for these clusters. Our detection of all 194 artificial
SS~Cygni in the field of M22 may be facilitated by its proximity, best
seasonal coverage and moderate core density (Fig.~26). The true
efficiency of our search in this case is probably close to 100\%. 

At this point it is prudent to discuss whether our search is
affected by stellar density and/or distance from the core centre.
On the one hand, we must remember that variable temporal sampling
yielded marked statistical variability of detection efficiency
(Fig.~25). On the other hand, as illustrated in Fig.~26, the image
subtraction is efficient in drastically reducing density of
features per image. Additionally, inspection of
Fig.~26 reveals that any stellar density gradient is not
apparent in the subtracted image, containing mostly
variable stars. A separate study of the top and bottom half of
images revealed no spatial efficiency effect comparable to the
temporal one. Effectively our detection algorithm is
applied to sparsely populated residual images and its 
efficiency is dominated by temporal coverage and to a lesser degree
to magnitude/distance effects. Exceptions were the tiny collapsed
cores of M30, NGC~362 and NGC~6752 where, due to saturation,
no $V$ photometry was obtained and no simulations were possible.

From histograms in Fig.~25 we conclude that the DNe detection
efficiency is dominated by a combination their outburst
duty cycle and the number of nights per cluster. The overwhelming
majority of larger duty cycle outbursts of SS~Cyg-type was
detected. For a fraction of smaller duty cycle of U~Gem-type
stars and less intensely observed clusters no outbursts
occurred during observations (the white areas in Fig.~25). For the
most sparsely observed cluster, NGC 2808, out of 250 U~Gem-type DNe
only 57 had observable outbursts and we recovered 42 of them. Thus
the effect of distance and magnitude is less pronounced, in the extreme
case of NGC 2808 yielding only 25\% missed dwarf novae in outburst
(sparsely hatched areas). Some DNe in outburst were not recovered at all.
They were too faint and scattered too much, or they were placed
too close to very bright stars to be detected. The radial
distributions of such objects are shown in Fig.~27. It is clear that
for clusters with sparse and non-saturated cores, like M22
(and here also M4, M5, M10, M12, M55, NGC~288, NGC~3201,
NGC~4372, NGC~6362, $\omega$~Centauri), completeness is independent
of the position in the cluster. However, for clusters with
small, dense and partly saturated cores one expects rather strong
radial dependence toward the centre.

We undertook additional simulations for the dense central part
of the globular cluster NGC~2808. Fig.~28 shows the area adopted
for the simulations. Inside a circle of radius $60\arcsec$
we inserted light curves of 1000 artificial SS~Cyg variables,
alongside 1000 artificial U~Gem variables. In this run all
simulated stars had outbursts in our data. We recovered 341 SS~Cyg
and 165 U~Gem stars. However, none of the stars was detected
closer than $8\farcs9$ from the cluster centre. Evidently
this is due to high number of saturated stars. The number of
recovered DNe increases with the distance from the centre,
as demonstrated in Fig.~29. The globular cluster NGC~2808
is the most distant of the all 16 clusters analysed in this work.
Therefore estimates of completness of the search for DNe
in this cluster can be treated as a worst case scenario
for the whole sample.

\begin{figure}
\vspace{8.3cm}
\includegraphics{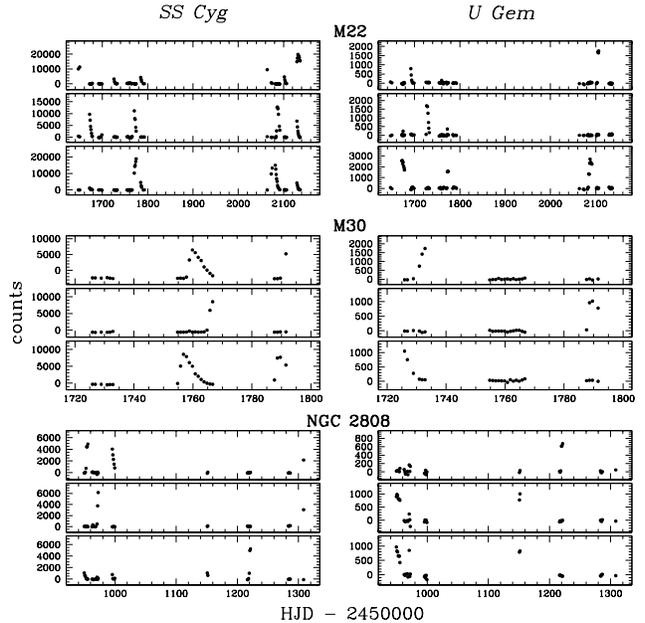}
\caption{\small Example light curves of artificial DNe inserted into
images of the globular clusters M22, M30 and NGC 2808. The clusters are
arranged from top to bottom by increasing distance modulus. Note that
negative counts (e.g. in the middle left panel) are a natural product
of the image subtraction method.}
\label{fig24}
\end{figure}

\begin{table*}
\begin{minipage}{100mm}
\begin{center}
\caption{Results of the simulations for the three clusters.}
{\small
\begin{tabular}{|l|c|c|c|c|c|c|}
\hline
& \multicolumn{2}{|c|}{M22} & \multicolumn{2}{|c|}{M30 without core} & \multicolumn{2}{|c|}{NGC 2808} \\
                      &  SS~Cyg &  U~Gem  &  SS~Cyg &  U~Gem  &  SS~Cyg &  U~Gem  \\
\hline
Number of       &         &         &         &         &         & \\
artificial DNe  &   194   &   193   &   250   &   250   &   250   &   250  \\
in the field    &         &         &         &         &         & \\
\hline
DNe in eruption &   194   &   140   &   217   &  ~~97   &   239   &  ~~57  \\
      & {\bf 100~\%} & {\bf 73~\%} & {\bf 87~\%} & {\bf 39~\%}  & {\bf 96~\%} & {\bf 23~\%} \\
\hline
Detected DNe    &   194   &   136   &   214   &  ~~91   &   238   &  ~~42  \\
      & {\bf 100~\%} & {\bf 69~\%} & {\bf 86~\%} & {\bf  36~\%} & {\bf 95~\%} & {\bf 17~\%} \\
\hline
\end{tabular}}
\end{center}
\end{minipage}
\end{table*}

\begin{figure}
\vspace{8.2cm}
\includegraphics{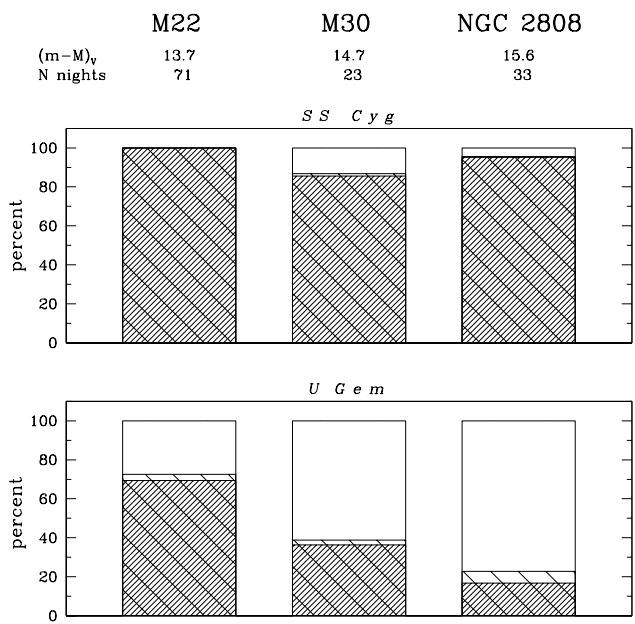}
\caption{\small 
Efficiency of detection of artificial DNe of SS~Cyg (upper panel)
and U~Gem type (lower panel) in sample globular clusters.
White areas correspond to stars whose eruptions missed 
our observations (the sampling/duty cycle effect). 
Densely hatched areas refer to the fraction of recovered DNe.
The remaining intermediate sparsely hatched area corresponds
to missed DNe which had outbursts during our observations
(the distance/magnitude effect). Note that the results do not apply
to the tiny collapsed core of M30, which is saturated on our frames.}
\label{fig25}
\end{figure}

\begin{figure*}
\vspace{13.0cm}
\includegraphics{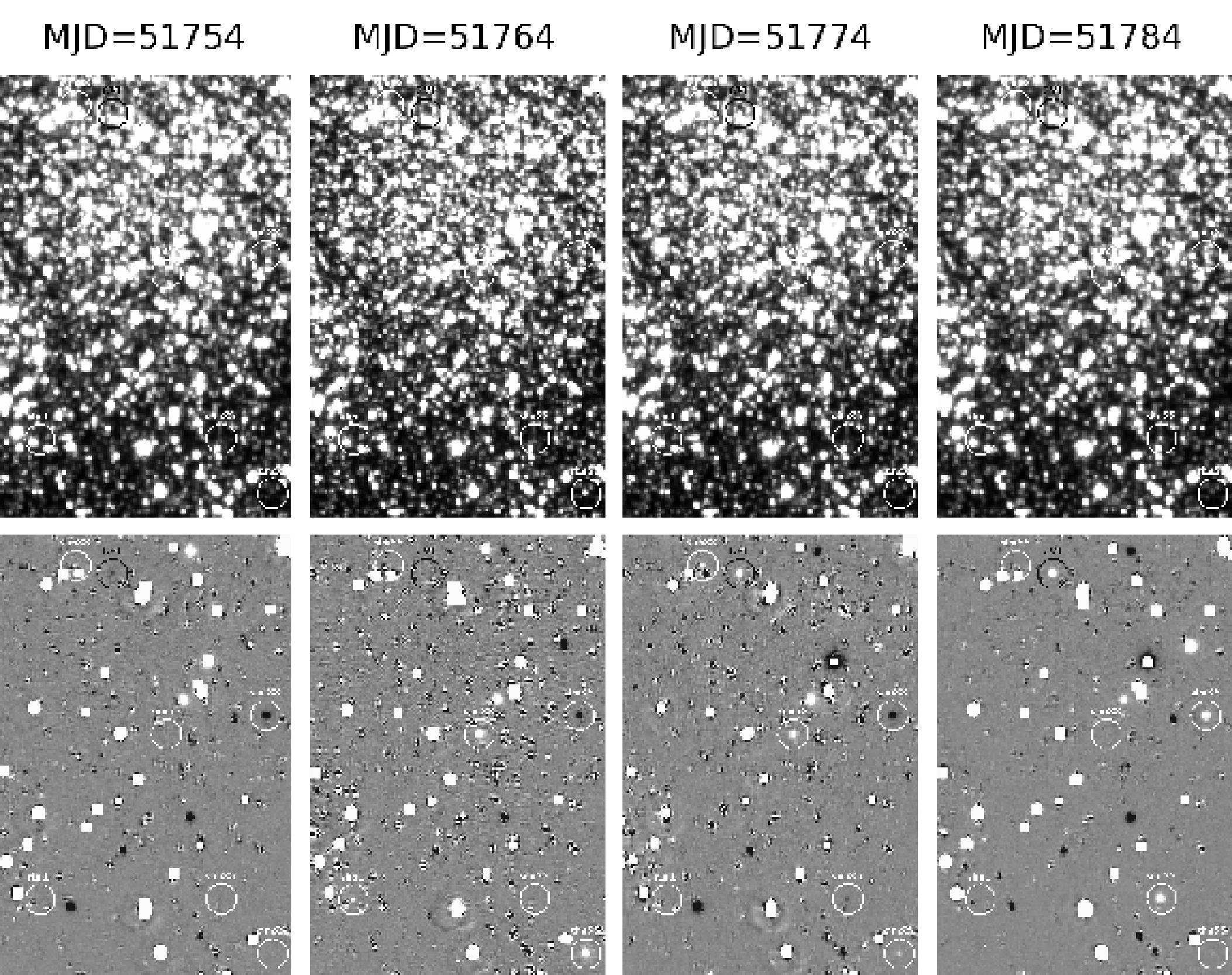}
\caption{\small Images of the same central region of the globular cluster M22
taken on four nights every 10 days in Aug/Sep 2000.
The area has $87\arcsec \times 130\farcs5$ ($200 \times 300$ pixels) and
is almost entirely located inside the cluster core, of radius
of $85\farcs2$. The lower panels show corresponding residual images.
The black circle denotes the known dwarf nova M22-CV1 while white circles denote
artificial dwarf novae inserted into frames of the cluster. The brightest
stars are saturated and are covered with white, rectangular areas.
The images demonstrate that in the image subtraction technique cores of
some globular clusters remain sparse fields as far as variable
stars are concerned.}
\label{fig26}
\end{figure*}

\begin{figure}
\vspace{4.3cm}
\includegraphics{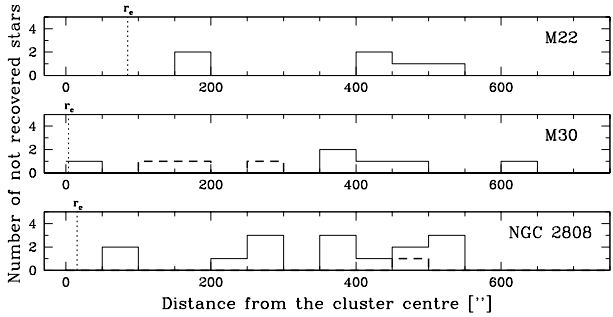}
\caption{\small Radial distribution of unrecovered artificial DNe
(only those whose eruptions occurred in our data) which were
inserted into the frames of the clusters M22, M30 and NGC 2808.
The thin solid lines refer to artificial U~Gem stars, while the thick
dashed lines refer to artificial SS~Cyg stars. Serendipitously, all synthetic
SS~Cyg stars in images of M22 were recovered. Core radii of the clusters
are marked with dotted lines. Note that our simulations do not apply
to the very small collapsed cores of M30.}
\label{fig27}
\end{figure}

\begin{figure}
\vspace{5.8cm}
\includegraphics{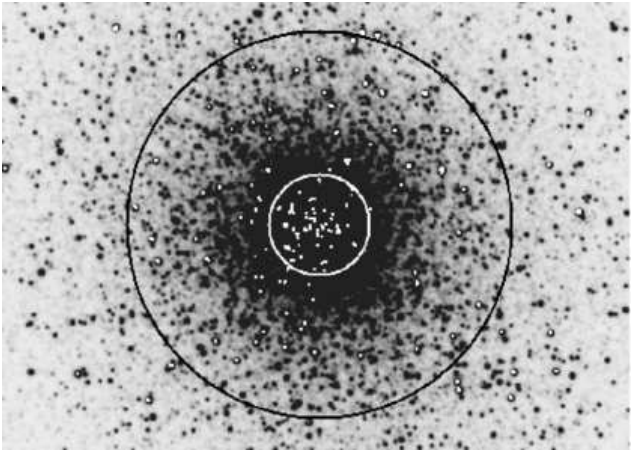}
\caption{\small Central part of the template image of the globular cluster
NGC 2808. North is up and East to the left. The large black circle of a
radius of $60\arcsec$ defines the area adopted for additional simulations.
The small white circle marks the cluster core (which radius is $15\farcs6$).
White patches in the image show saturated areas. Note that sizes of the areas
are smaller or larger in other images, depending on exposure time and seeing.
}
\label{fig28}
\end{figure}

\begin{figure}
\vspace{5.6cm}
\includegraphics{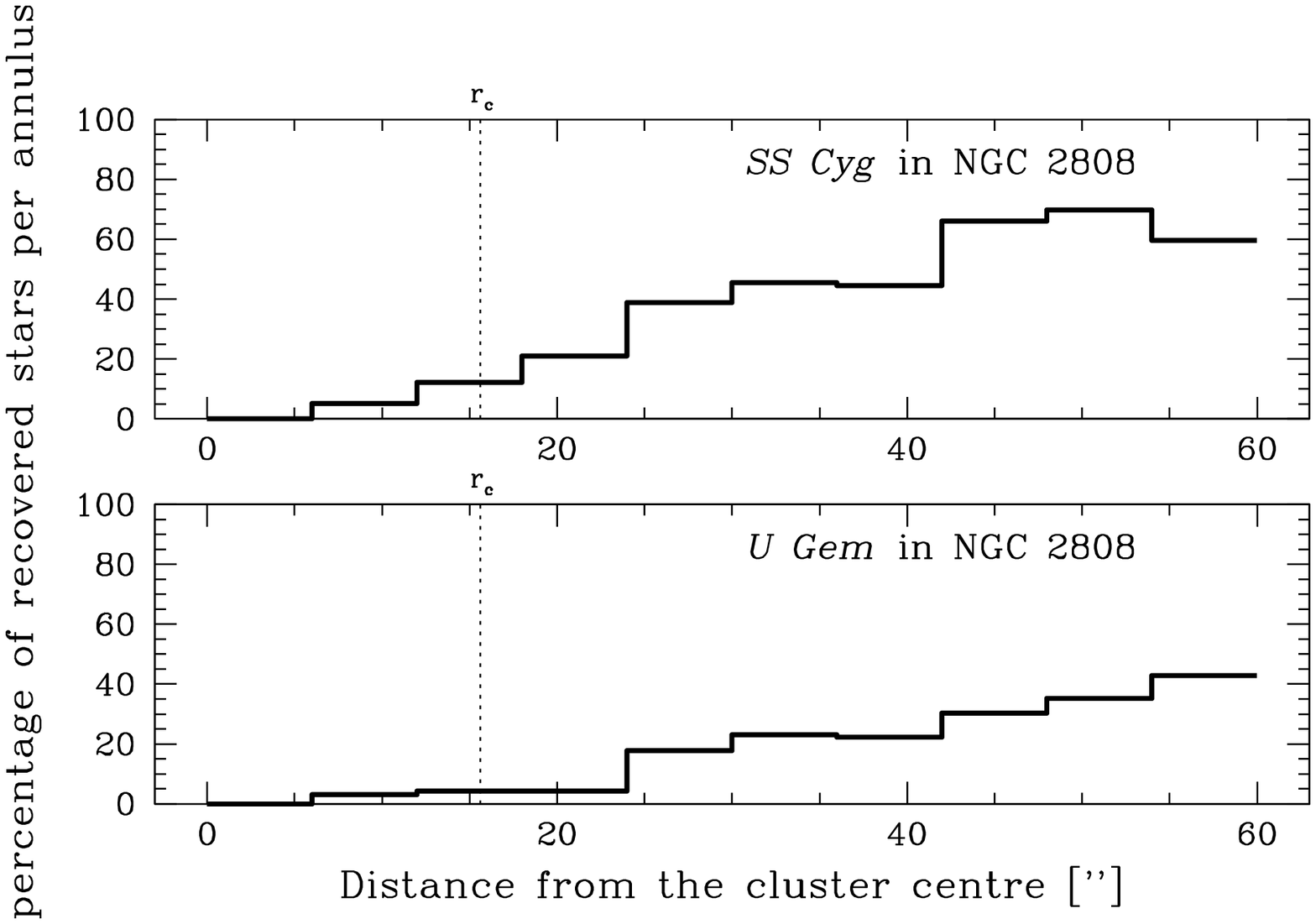}
\caption{\small Radial distributions of recovered artificial DNe
in additional simulations performed for the central part of the
globular cluster NGC 2808, the most distant globular cluster in our
sample. Each bin (annulus) is $6\arcsec$ wide.
The core radius of the cluster is marked with dotted lines. It can be
seen that the number of recovered stars clearly decreases toward
the centre. It is also evident that fainter U~Gem stars are harder
to detect than brighter SS~Cyg stars.
}
\label{fig29}
\end{figure}

Our simulations for the clusters M22, M30 and NGC~2808
allowed us direct estimation of the magnitude range of our DNe
search. For these clusters we determined the limiting magnitude,
corresponding to the faintest
recovered artificial star in outbursts. For the remaining clusters
the limiting magnitudes $m_{\rm c}$ were extrapolated from these
results using the following formula: $$ m_{\rm c}=m_{\rm
M22}-2.5~log ~\frac{\langle t_{\rm M22} \rangle}{\langle t_{\rm c}
\rangle} ~\frac{n_{\rm c}}{n_{\rm M22}}, $$ where $m_{\rm M22}$ is
the limiting magnitude for M22, $\langle t_{\rm M22} \rangle$ and
$\langle t_{\rm c} \rangle$ are the mean exposure times of the frames,
while $n_{\rm M22}$ and $n_{\rm c}$ are the median background
levels for M22 and a given cluster, respectively. These limiting
magnitudes are listed in Table~3.


\section{Discussion}

Our survey of DNe globular clusters is based on the largest
available homogeneous sample of observations, in terms of the
time span of several years, number of observations and number of clusters.
It extended over 16 Galactic GCs and yielded two new certain DNe:
M55-CV1 and M22-CV2. All previously known systems located in
our fields were recovered, too. For 16 GCs our survey yielded on
average 50 nights per cluster, much more than any
survey conducted so far. Past surveys of M92, M15 and of NGC~6712
extended over 10 nights \citep[independently:][]{sha94,tua03},
13 and 11 nights \citep{tua03},
respectively. In total there are 12 known DNe in 7 Galactic GCs.
Unfortunately, they constitute an insufficient sample to establish
general properties of the cluster DNe. Half of the objects are
located inside cluster cores (see Fig.~30; c.f., distances
listed in Table~1), and some systems are located several core radii
away from it.

Theoretical considerations predict that most CVs are concentrated inside
cluster cores \citep[see e.g.][]{iva06}. However in the crowded
cluster cores many detection methods suffer from strong observational
biases against CV detection. Due to use of the image subtraction,
our search was little affected by image crowding. Exceptional cases
were the tiny collapsed cores of M30, NGC~362 and NGC~6752,
and also dense and partly saturated core of NGC~2808
(the most distant cluster in our sample). In the first three
cases the cores cover from 30 to 300 seeing disks.
They are mostly saturated on our deep $V$-band exposures,
effectively yielding no data. This deficiency
was partly compensated by our outburst search on $B$-band
images of the core of NGC~362 and NGC~2808. In the $B$
band red giants do not saturate the core image and blue DNe outbursts
would easily be detected, but it must be stressed here that already in
the immediate vicinity of the saturated cores our photometry
performs well in the range of magnitudes discussed here and we
detect eclipsing and pulsating stars, reported elsewhere. In particular,
just several PSF diameters away from the saturated collapsed core
of NGC~362 we found a pulsating star of $V$ amplitude $0.3$ and
$M_V\approx3.5$ mag. This experience from our non-CV work
enhances our confidence that our detection efficiency does not
depend very much on the distance from the cluster centre.

Searches for quiescent cataclysmic variables in globular clusters
are also hampered due to the intrinsic faintness of these binaries.
The DNe detected in GCs so far have absolute magnitudes $M_V$
in quiescence ranging from +5.4 to +9.8~mag, or fainter.
Our simulations demonstrate that outbursts of amplitudes between
2 and 4 magnitudes yield improvement of their detectability also
in distant clusters, albeit strongly dependent on DNe duty cycle.
This may explain why all seven clusters harbouring 12 known
DNe are relatively nearby, less than $\sim$10 kpc away, and
well-covered by observations. In this context we emphasise that
we detected no light variation near the locations of 56 X-ray sources and
their optical counterparts, 27 of which are CV candidates.

\begin{figure}
\vspace{3.6cm}
\includegraphics{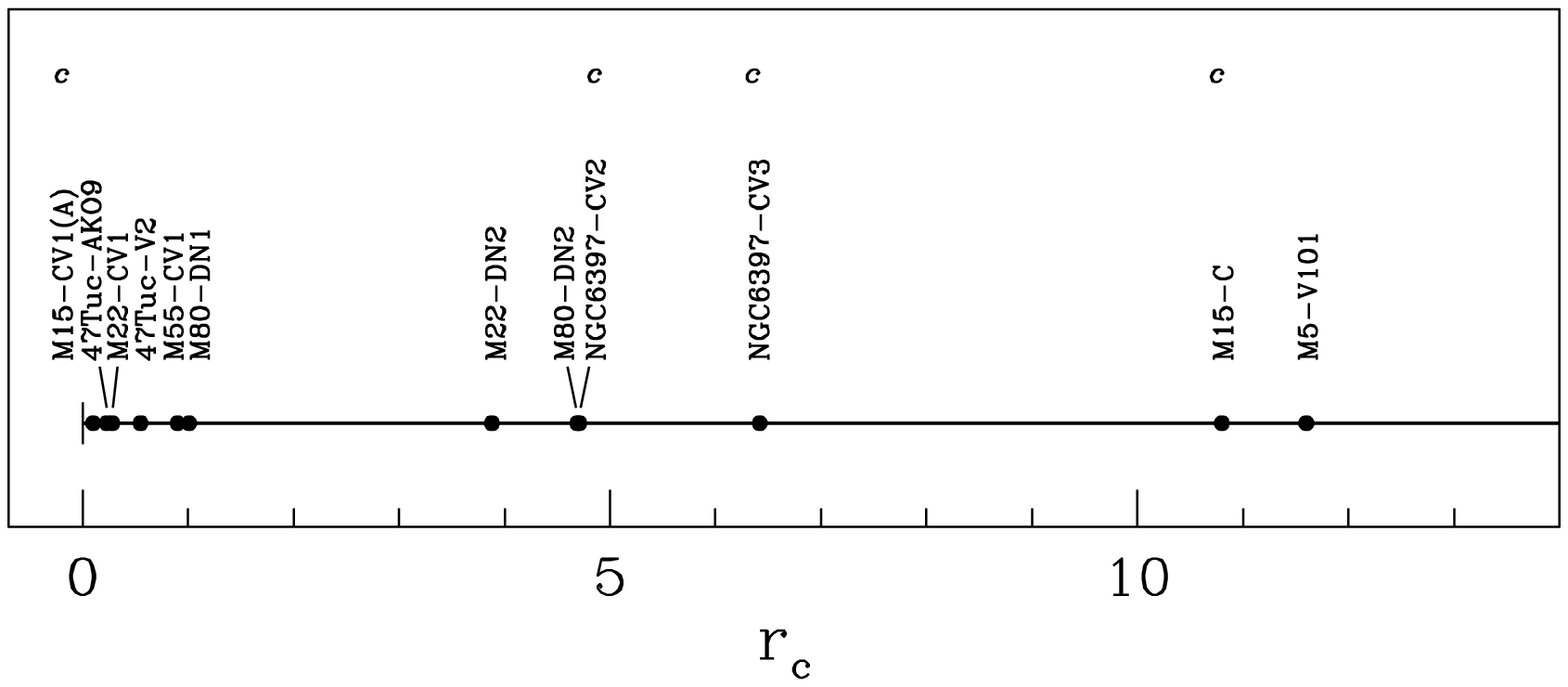}
\caption{\small Distances for 12 known GC DNe in units of core radii
from the cluster centres. Small characters '{\it c}' denote DNe
in core-collapsed clusters.}
\label{fig30}
\end{figure}

The environment of DNe in GC depends on cluster age and metallicity.
In Fig.~31 we plot these two parameters for 83 Galactic GCs
\citep[data taken from:][]{sal02,gra03,carr04,san04}, of which 7
contain DNe. Ages and metallicities for the seven clusters are
given in Table~6. Inspection of Fig.~31 reveals that the clusters
with DNe are not concentrated in any particular location in the
plot, they appear to be scattered among all clusters. Metallicity
of four out of five clusters harbouring 2 DNe, namely NGC 6397,
M15, M22 and M80, is low. This statistic is insufficient to show
any tendency. Such a tendency would contradict the presence
of numerous field DNe in the metal rich population I.

\begin{table}
\begin{center}
\caption{Metallicites (using Carretta \& Gratton 1997 scale)
and ages for globular clusters harbouring dwarf novae.}
{\small
\begin{tabular}{|l|c|c|c|}
\hline
Cluster  &     [Fe/H]     &     Age      & Source of data \\
         &                &    [Gyr]     & \\
\hline
M5       & $-1.11\pm0.03$ & $10.9\pm1.1$ & \citet{san04} \\
M15      & $-2.02\pm0.04$ & $12.9\pm0.6$ & \citet{san04} \\
M22      & $-1.68\pm0.15$ & $12.3\pm1.2$ & \citet{san04} \\
M55      & $-1.54\pm0.10$ & $12.3\pm1.7$ & \citet{sal02} \\
M80      & $-1.47\pm0.04$ & $13.7\pm0.9$ & \citet{san04} \\
NGC 6397 & $-2.03\pm0.05$ & $13.9\pm1.1$ & \citet{gra03} \\
47 Tuc   & $-0.67\pm0.05$ & $11.2\pm1.1$ & \citet{carr04} \\
\hline
\end{tabular}}
\end{center}
\end{table}

\begin{figure}
\vspace{8.3cm} \includegraphics{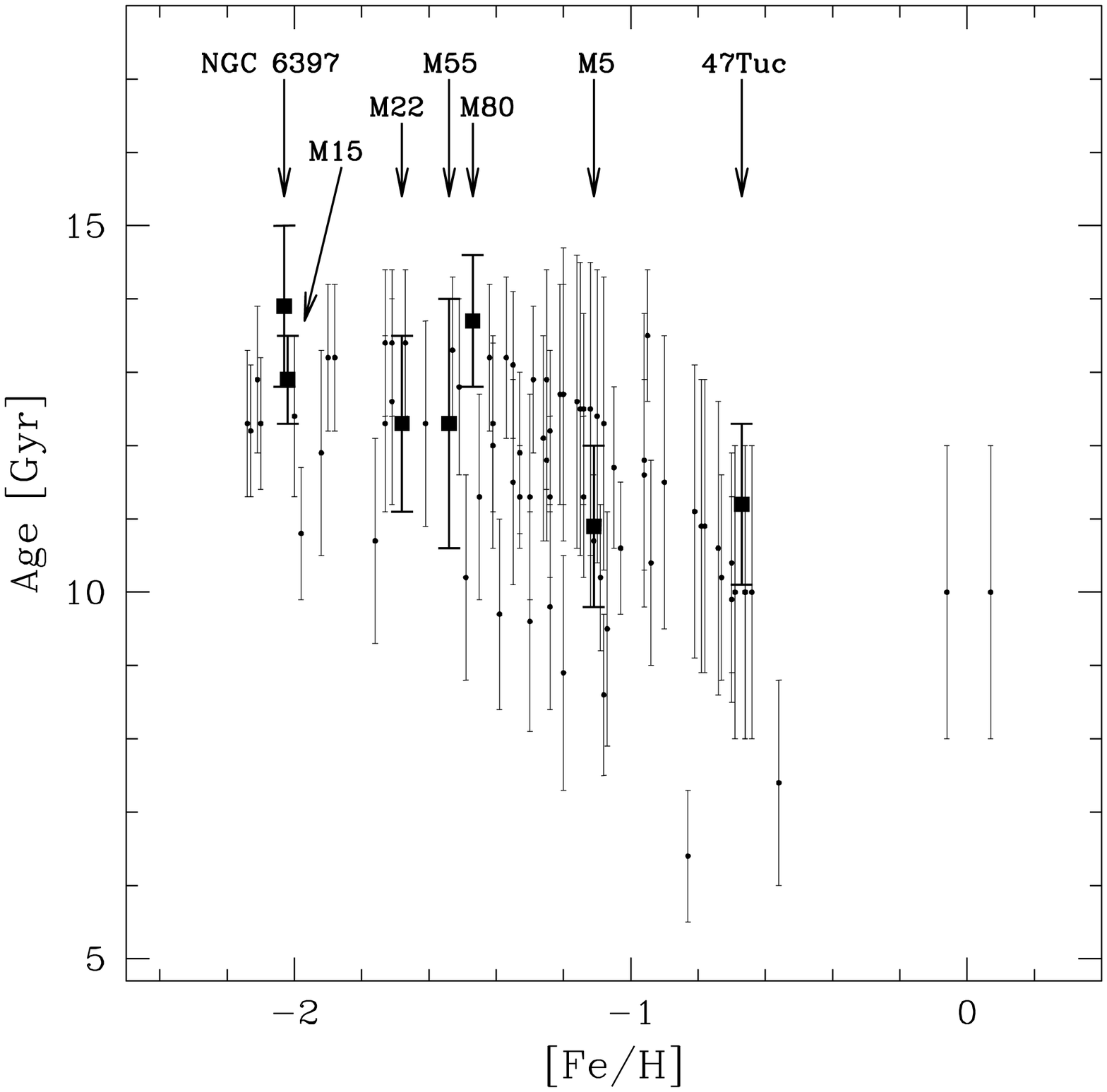} \caption{\small Age vs.
metallicity dependence for 83 Galactic globular clusters, 7 of
which have observed DNe. Metallicity error bars are not
plotted. The data for all the clusters (except NGC 6397 and 47~Tuc)
come from \citet{sal02} and \citet{san04}. The ages and metallicites
for NGC 6397 and 47~Tuc were taken from \citet{gra03} and
\citet{carr04}, respectively.}
\label{fig31}
\end{figure}

In Table~4 we list outburst properties of prototype DNe, scaled
according to cluster distances. Our simulations have confirmed
that if globular clusters were populated by numerous dwarf novae
resembling U~Gem or SS~Cyg they would be easily detected in our
survey. Detection efficiency of general DNe in our simulations
mostly depends on the duty cycle and observation coverage. The duty
cycle is not particularly discriminating as many, perhaps most
DNe have duty cycles comparable to these of U~Gem or SS~Cyg. On
the contrary, the outburst absolute magnitude is correlated with
the orbital period \citep{war95}. Thus our survey is slightly
biased towards detection of brighter, U~Gem-type, DNe above
the period gap, and against SU~UMa-type below it. For the most likely
periods (distribution modes) of these types (respectively 3.8 and
1.5 hours) the average difference of absolute magnitudes in
outburst amounts to roughly one magnitude. The corresponding loss in
detection efficiency of SU~UMa-type stars would correspond to
a shift by one panel to the right in Fig.~25, amounting to no more than 
25\% loss of observable outbursts. Paradoxically SU~UMa by itself would be
discovered as easily as U~Gem, because of a similar absolute brightness
and (super)outburst duty cycle. About a third of SU~UMa-type stars, i.e.,
WZ Sge stars, undergo extreme superoutbursts with large amplitude
and recurrence times longer than a year. We would miss most of
these superoutbursts because of their small duty cycle. However,
as observed by \citet{war95}, statistics of field CVs suffers from
unknown and possibly large selection effects, rendering detailed
quantitative comparison uncertain.

We have checked the literature to count the number of erupting CVs
in the solar neighbourhood. Out of 11 CVs located within the sphere
of a radius of 100~pc from the Sun, seven stars (AY~Lyr, EX~Hya,
U~Gem, V436~Cen, VY~Aqr, VW~Hyi, WZ~Sge) undergo outbursts
(the four remaining CVs are AM~Her, GP~Com, IX~Vel, V2051~Oph).
Within a 200~pc sphere we have counted 22 erupting cataclysmic
variables (DI~UMa, GW~Lib, HT~Cas, OY~Car, SS~Aur, SS~Cyg, SW~UMa,
SX~LMi, T~Leo, V893~Sco, WW~Cet, WX~Hyi, YZ~Cnc, Z~Cam, Z~Cha,
plus seven variables mentioned above) out of 45 objects
(the remaining stars are AE~Aqr, AH~Eri, BK~Lyn, BL~Hyi, BY~Cam,
EF~Eri, FO~Aqr, HU~Aqr, KR~Aur, MR~Ser, RW~Sex, ST~LMi, TT~Ari,
V347~Pav, V834~Cen, V3885~Sgr, VV~Pup, WX~Ari, YY~Dra).
These numbers show that about half of the known CVs exhibit
outbursts. However one should remember that distances for
a significant number of known field CVs are uncertain
or have never been measured. Based on data from the electronic
edition of the Catalogue of Cataclysmic Binaries,
Low-Mass X-Ray Binaries and Related Objects
\citep[][RKcat Edition 7.9, Jan 1, 2008]{rit03}
we find median recurrence times of 101 and 49 days
for erupting CVs with distances up to 100~pc and 200~pc
from the Sun, respectively. 

One can roughly calculate the number of known CVs per unit stellar
mass in the solar vicinity. If we take the value of
local mass density of 0.035 $M_{\sun}/pc^3$
\citep[0.031 $M_{\sun}/pc^3$ for the main sequence stars
plus 0.004 $M_{\sun}/pc^3$ for white dwarfs, respectively from][]
{reid02,reid97}, we find that the 100~pc sphere contains
$\sim1.5 \times 10^5M_{\sun}$, and the 200~pc sphere
about $1.2 \times 10^6M_{\sun}$. This gives approximately
$4.8\times 10^{-5}$ known erupting CVs per
$M_{\sun}$ in the first case and $\sim1.9\times 10^{-5}$
CVs per $M_{\sun}$ in the second one.
For comparison, the estimated mass of the globular cluster
M22 is $2.7 \times 10^5M_{\sun}$ \citep{alb02},
47~Tuc is of $1.3 \times 10^6M_{\sun}$ \citep{mey86}, and
$\omega$~Cen of $2.9 \times 10^6M_{\sun}$ \citep{mey86}.
This translates into $\sim7.4 \times 10^{-6}$,
$1.5 \times 10^{-6}$ and 0 known DNe per unit solar mass
respectively in the three clusters. All these densities
for the clusters are significantly smaller than those
estimated for the solar neighbourhood.

While it seems reasonable to conclude that, compared to the solar
vicinity, GCs are deficient in U~Gem-like variables,
i.e., dwarf novae with recurrence times up to a few hundreds
of days, the actual cause remains obscured.
Either the relative number of DNe is small or their outburst
properties are different. Rare outbursts in GC CVs would
correspond to small average transfer rates and dim quiescent
accretion discs compared to those in CVs in our vicinity
(\citealt{war95}, his Fig.~3.9 and Eq. 3.3). {\it HST} observations of
low optical fluxes from 22 CV candidates would be consistent with
such a hypothesis \citep{edm03}. An alternative explanation
postulates that most CVs in GCs are magnetic and avoid
outbursts as normal intermediate polars do. In these stars the
magnetic field of the white dwarf truncates the inner accretion disc
thus preventing or diminishing its outbursts. According to this
proposal, in clusters close encounters or even stellar mergers
result in a faster rotation of stellar cores and thus they facilitate
growth of the magnetic field by stellar dynamo mechanism
\citep{iva06}. The detection of relatively strong He\textsc{ii}
emission of CV1, CV2 and CV3 in NGC 6397 would be consistent with
their magnetic nature \citep{grin95}. \citet{edm99} showed that
CV4 in NGC 6397 resembles intermediate polars. Observations of
outbursts of CV2 and CV3 by \citet{sha05} and \citet{kal06} do
not invalidate the magnetic hypothesis as similar outbursts were
observed in confirmed intermediate polars (e.g. GK Per, EX Hya).
Further credit to the magnetic hypothesis was lent by the detection of
intermediate polar-like 218~s oscillations in X-ray source X9 in
47~Tuc \citep{hei05}. Another X-ray source in the same cluster, X10,
exhibits a 4.7~hour period and an X-ray spectrum consistent with those
seen in polars. 

\citet{dob06} proposed that rare outbursts
in some CVs result from the combination of low mass transfer rates
($10^{-13}-10^{-12}$ $M_{\sun}$/year) and moderately strong white
dwarf magnetic moments ($>10^{30}$ G cm$^3$). Conversely,
the mass transfer rates in GC CVs obtained in numerical
simulations by \cite{iva06} exceed those proposed
by \citet{dob06} by two orders of magnitude. According to
\citet{iva06}, a field strength of $10^7$ G suffices to prevent DN
outbursts in all CVs with the white dwarf mass less than 1.1
$M_{\sun}$.

In an alternative scenario, frequent stellar encounters affect
stability of binary orbits, thus affecting their ability to sustain
an accretion rate suitable for formation of an accretion disc.
Recently, \citet{sha06} simulated a 100,000 star cluster with 5000
primordial binaries. They found that cluster CVs were affected
by encounters with other cluster members. This
tends to shorten their life and hence decrease the expected
number of active CVs, at any given time, by a factor of 3,
compared to field CVs.

Extensive observations of the known CV candidates in clusters are
urgently needed to determine their average accretion rates and
dominant mode of variability. In particular, it would be important
to verify whether these objects are magnetic. Magnetic fields in
such stars are reliably detected by Zeeman split of the lines,
strong and/or variable polarisation and by presence of the
cyclotron harmonics (humps) in their continuum spectra.


\section{Summary}

We report the results of our extensive photometric survey by the
CASE collaboration of 16 Galactic GCs with the aim of detecting DN
outbursts. Our early discovery of two DNe, namely CV1 in the
globular cluster M55 and CV2 in M22 were reported in \citet{kal05a}
and in \citet{pie05}, respectively, where we find
that the former object is likely to be a U~Gem-type DN,
while the latter one is an SU~UMa DN, and their X-ray counterparts
were detected by the XMM-Newton satellite. Their positions in the
color-magnitude diagrams of the clusters and their X-ray fluxes
are consistent with their cluster membership.

In the present paper we report that a similar search of DNe in the
remaining 14 GCs yielded no new DNe. Our search was sufficiently
sensitive to easily detected outbursts of all three well-known DNe
in our fields: V101 in the globular cluster M5, CV1 in the core of
M22, and the DN V4 in the foreground of the globular cluster M30.
We investigated further the nature of the enigmatic erupting
source NV408, located $\sim2\arcmin$ from the centre of
$\omega$~Centauri, but found no evidence confirming its CV nature.

It is remarkable that we detected no outbursts nor any detectable
variability near positions of 27 known CV candidates. To verify
the efficiency of our survey we performed simulations by inserting
into real frames artificial stars with light curves mimicking
those of SS~Cygni and U~Geminorum-type DNe. The results for M22,
M30, and NGC 2808 demonstrated we were able to recover between
16-100\% of artificial DNe, depending on their duty cycle 
and observation coverage. For larger duty cycle SS~Cyg-type stars
our coverage was sufficient to obtain very close to 100\% recovery rate.
However, some U~Gem type stars with a smaller duty cycle had
no outbursts on our nights and hence for extremely poor sampling
up to 70\% of them were missed. Any effects of stellar
crowding were diminished by the application of the image
subtraction technique for relatively sparse cores of some clusters,
like M22, M55 or $\omega$~Centauri. For clusters with very dense
(and usually partly saturated) cores, the number of recovered stars
decreases toward the centre, as was demonstrated in additional
simulations for the central part of NGC 2808.

The results of our extensive survey provide new evidence
confirming early suggestions, based on more fragmentary data, that
ordinary DNe are indeed very rare in GCs. On one hand, up to 150
candidate CVs were proposed in GCs, mostly from X-ray surveys. On
the other hand our survey combined with earlier results yielded
only 12 confirmed DNe in total in the substantial fraction of all
Galactic globular clusters. Such a fraction of DNe among CVs
appears extremely low by comparison to the field CVs, of which
half are DNe. Thus our observations acutely
question why outbursts of dwarf novae are rare in GCs. We quote
some recent evidence that most cluster CVs could be magnetic and
thus exhibit little or no outbursts. However, any conclusions must
await a thorough study of all cluster CV candidates.


\section*{Acknowledgments}

We acknowledge with thanks the variable star observations
from the AAVSO International Database contributed by observers
worldwide and used in this research. Some of the data
presented in this paper were obtained from the Multimission
Archive at the Space Telescope Science Institute (MAST),
which is operated by the Association of Universities for
Research in Astronomy, Inc., under NASA contract NAS 5-26555.
The observations are associated with the programme \#9442.
PP was supported by Polish MNiI grant number N203 019 31/2874
and the Copernicus Foundation for Polish Astronomy.
PP and JK acknowledge support from the Domestic Grant for Young
Scientists and from the Grant MASTER of the Foundation
for Polish Science, respectively. The CASE project is
supported by the NSF grant AST 05-07325 and by Polish
MNiI grant 1P03D 001 28. It is also a pleasure to thank
Michael Shara and Sophia Khan for remarks on the draft
version of this paper. We are grateful to the referee for
very quick and detailed report on the paper.

\end{document}